\documentstyle[floats,prd,aps,epsf]{revtex}

\begin{document}
\def\go{\mathrel{\raise.3ex\hbox{$>$}\mkern-14mu\lower0.6ex\hbox{$\sim$}}}
\def\lo{\mathrel{\raise.3ex\hbox{$<$}\mkern-14mu\lower0.6ex\hbox{$\sim$}}}
\newcommand{\bzeta}{\mbox{\boldmath $\zeta$}}
\newcommand{\ratmas}{M_{{\mathit NS}}/M_{{\mathit BH}}}
\newcommand{\VC}{{\mathit VC}}
\newcommand{\ratmasinv}{M_{{\mathit BH}}/M_{{\mathit S}}}
\newcommand{\st}{{\mathit S}}
\newcommand{\ns}{{\mathit NS}}
\newcommand{\bh}{{\mathit BH}}
\newcommand{\gr}{{\mathit GR}}
\newcommand{\inin}{{\mathrm in}}
\newcommand{\ouou}{{\mathrm out}}
\newcommand{\Jseq}{J(d_G)}
\newcommand{\Eseq}{E(d_G)}
\newcommand{\bJseq}{\bar{J}(\bar{d}_G)}
\newcommand{\bEseq}{\bar{E}(\bar{d}_G)}
\newcommand{\bOseq}{\bar{\Omega}(\bar{J})}
\newcommand{\dg}{d_G}
\newcommand{\rse}{r_{{\mathit sec}}}
\newcommand{\rdy}{r_{{\mathit dyn}}}
\newcommand{\rR}{r_R}
\newcommand{\bdg}{\bar{d}_G}
\newcommand{\brse}{\bar{r}_{{\mathit sec}}}
\newcommand{\brdy}{\bar{r}_{{\mathit dyn}}}
\newcommand{\brR}{\bar{r}_R}
\newcommand{\tX}{{\widetilde{X}}}
\newcommand{\tY}{{\widetilde{Y}}}
\newcommand{\tZ}{{\widetilde{Z}}}
\newcommand{\fnabla}{\,^{(4)}\nabla}
\newcommand{\beq}{\begin{equation}}
\newcommand{\eeq}{\end{equation}}
\newcommand{\beqn}{\begin{eqnarray}}
\newcommand{\eeqn}{\end{eqnarray}}
\newcommand{\pa}{\partial}

%
%
\twocolumn[\hsize\textwidth\columnwidth\hsize\csname
@twocolumnfalse\endcsname

\title{A new numerical method for constructing quasi-equilibrium sequences 
of irrotational binary neutron stars in general relativity}

\author{K\=oji Ury\=u}
\address{SISSA, Via Beirut 2/4, Trieste 34013, Italy \\
International Center for Theoretical Physics, 
Strada Costiera 11, Trieste 34100, Italy}
\author{Yoshiharu Eriguchi}
\address{Department of Earth Science and Astronomy,
Graduate School of Arts and Sciences,
University of Tokyo, Komaba, Meguro, Tokyo 153-8902, Japan}
\maketitle
\begin{abstract}
We propose a new numerical method to compute quasi-equilibrium sequences of 
general relativistic irrotational binary neutron star systems.  It is a good 
approximation to assume that
(1) the binary star system is irrotational, i.e. the vorticity of the 
flow field inside component stars vanishes everywhere (irrotational flow), 
and  
(2) the binary star system is in quasi-equilibrium, 
for an inspiraling binary neutron star system 
just before the coalescence as a result of gravitational wave emission. 
We can introduce the velocity potential for such an irrotational flow field, 
which satisfies an elliptic partial differential equation (PDE) with a 
Neumann type boundary condition at the stellar surface.  For a treatment of 
general relativistic gravity, we use the Wilson--Mathews formulation, which 
assumes conformal flatness for spatial components of metric.  
In this formulation, the basic equations are expressed by a system of elliptic 
PDEs.  We have developed a method to solve these PDEs with appropriate 
boundary conditions.  The method is based on the established prescription for 
computing equilibrium states of rapidly rotating axisymmetric neutron stars or 
Newtonian binary systems.  We have checked the reliability of our new code by 
comparing our results with those of other computations available.  We have 
also performed several convergence tests.  By using this code,
we have obtained quasi-equilibrium sequences of irrotational binary star 
systems with strong gravity as models for final states 
of real evolution of binary neutron star systems just before coalescence.  
Analysis of our quasi-equilibrium sequences of binary star 
systems shows that the systems may not suffer from dynamical instability of 
the orbital motion and that the maximum density does not increase 
as the binary separation decreases. 

\end{abstract}
\vskip2pc]   
%

\section{Introduction}\label{intro}

At the beginning of the next century, advanced gravitational wave 
detectors (LIGO/VIRGO/TAMA/GEO; \cite{EXPREV}) will provide us with 
signals from coalescing binary neutron star systems.  According to the 
type of compact objects involved (for example, black holes/neutron 
stars, isolated/binary objects, solar mass/super massive objects),  
various theoretical methods have been developed or are under development 
to predict and interpret gravitational wave signals which will be 
observed.  Compact binary systems of solar mass sizes 
are also promising sources of $\gamma-$ ray bursts or r-process nuclei. 
These are important issues of recent astrophysics (see for example 
\cite{BNSREV} and references therein).  

When we consider the final evolutionary stage of the binary neutron 
star (BNS) as a result of gravitational wave emission, 
it is necessary to solve the internal structure 
of the neutron star (NS) at the late inspiraling stage 
as well as the early stage of merging, 
since the size of the star is comparable to the 
separation of the two neutron stars.  
In this case, (1) tidal deformation of a component star, 
(2) the general relativistic effect on the orbital motion, 
and (3) the general relativistic effect on the internal structure of a star
are coupled to affect the physics of the binary system.  
Such finite size effects are not negligible for the last 
${\cal N} \lo 10$ cycles of the inspiraling phase just before 
coalescence \cite{BNSREV}.  

According to the scenario of the final evolution of the close 
BNS \cite{kbc92}, the flow field inside component stars just prior 
to coalescence becomes {\it irrotational}, i.e. a vorticity seen from 
the non-rotating observer's frame vanishes just prior to coalescence.  
This irrotational flow is realized because (1) the viscosity of neutron 
star matter is too weak to synchronize the spins of the component stars 
and the orbital angular 
velocity of the binary system during the evolution, 
and (2) the initial spin period of each NS is much larger than the final 
orbital period of the BNS.  It is on precisely this stage of the binary 
evolution that we focus in the present paper.  

Full 3D numerical simulations are most promising to investigate such final 
phases of BNS systems.  Numerical methods for fully dynamical 3D GR 
computations of BNS's are developing rapidly (see for example 
\cite{on97,ss99,sh99}).  
However it is still difficult to perform simulations over the whole 
${\cal N} \lo 10$ cycles of the inspiraling BNS because 
there remain some technical problems to overcome for newly developed 
3D GR codes, such as, 
maintaining stability and accuracy of numerical schemes and 
restrictions on the computational power of present computers.  
Recently, Shibata has succeeded in developing a new numerical method 
to compute dynamical evolution of the space-time and compact 
object(s) such as a single NS or a BNS system in a stable manner \cite{sh99}.  
Therefore simulations over several orbital periods of BNS will be performed 
in the near future.  However, 
even after that, it is desirable to have totally different 
and supplementary methods to investigate such systems to test 
the reliability of the theory based on numerical methods.  

One of promising alternative ways to tackle this problem is to compute 
quasi-equilibrium sequences of BNS configurations.  Since the time scale 
of evolution as a result of gravitational wave (GW) emission is longer 
than the time scale of the orbital motion even at this stage, 
quasi-equilibrium 
is a good approximation for a realistic BNS system \cite{st83}.  
Therefore, a quasi-equilibrium sequence with a constant rest mass 
constructed by changing the binary separation could be 
considered as an evolutionary sequence of such an inspiraling BNS system.  
Also, quasi-equilibrium configurations will give accurate initial 
conditions for the future 3D GR dynamical computations.  Therefore, 
it is important to develop a numerical method to compute 
quasi-equilibrium configurations of close BNS's.  

In this paper we introduce a new numerical method to compute 
quasi-equilibrium configurations of irrotational BNS systems.  
It is a straightforward extension of 
the computational method used for the Newtonian irrotational binary 
systems developed by the present authors \cite{ue98}.  
We discuss the formulation of the problem separately for 
(1) the gravitation part (the Einstein equations) and 
(2) the fluid part (the NS, i.e., the equation of motion, the continuity 
equation, the equation of state (EOS) and so on).  
As the first step of the extension to general relativistic strong 
gravity, a simplified system of the Einstein equations is employed.  
We use the formulation developed by 
Wilson and Mathews \cite{wm956} and Baumgarte et al. 
\cite{ba97}, 
in which the metric is decomposed into (3+1) form and 
its spatial part is assumed to be conformally flat. 
For the fluid part, we solve the equations for irrotational 
flow whose formulation in GR has been developed by Shibata \cite{sh98} 
and Teukolsky \cite{te98} (see also \cite{bas978}).  
Our new computational code reproduces results of several independent 
computations previously made for GR co-rotating BNS's and Newtonian 
irrotational binary systems with reasonable accuracy.  
The convergence test of the numerical scheme is shown as well. 
Using this new code, 
we have constructed quasi-equilibrium sequences of irrotational BNS's 
with constant rest mass, which approximate the final evolutionary 
track of a realistic BNS as a result of GW emission.  
Analysis of our quasi-equilibrium sequences of binary star systems 
suggests that the systems may not suffer from dynamical instability of 
the orbital motion and that the maximum density does not increase 
as the binary separation decreases. 

This paper is organized as follows.  In section II we review the 
formulation for quasi-equilibrium irrotational binary systems 
in GR and derive the basic equations.  In section III, we present 
the new numerical method for solving them.  In section IV, 
we present several comparisons with other computations, 
a convergence test of the new numerical scheme and the results for 
the binary configurations and quasi-equilibrium sequences.  
In section V, we discuss the validity of 
the present results and future prospects.  

\section{Review of the formulation of the problem}

\subsection{Quasi-equilibrium irrotational binary systems in GR: 
Summary of the assumptions and of the formulation} \label{assump}

Because of the finite size effect of stars in a close 
binary system, it is necessary to solve for the structures of the stars 
without approximation for the last ${\cal N} \lo 10$ cycles of 
the inspiraling phase just before coalescence.  
In general, the binary system cannot be in equilibrium in GR 
since GW carry angular momentum and energy away to infinity.  
However, the quasi-equilibrium approximation is well fulfilled 
since the time scale of evolution due to 
GW emission is longer than that of the orbital period 
even just before coalescence \cite{st83}.  
In this situation, we can expect that (A) there exists 
an approximate Killing vector $\vec{\ell}$ expressing 
this symmetry of the space-time and the fluid. 
Moreover, it has been pointed out by Kochanek and by Bildsten \& Cutler
\cite{kbc92} that (B) the flow fields of binary stars just before 
coalescence become {\it irrotational}, because (i) the viscosity of 
the neutron star matter is too weak to synchronize the spins and 
the orbital angular velocity by the time of merging and 
(ii) the initial spin angular velocity of each component 
star is negligible compared to the final orbital angular velocity.  

To construct realistic BNS's in quasi-equilibrium, 
the above two properties (A) and (B) should be implemented in the 
formulation of the basic equations.  
Strictly speaking, it is impossible to assume the property 
(A) in exact GR, because there arises GW emission from the source to 
the asymptotically flat region.  This implies that there is no unique 
way to formulate properly an approximation which fulfills 
the property (A) for the Einstein equations.  
However, several authors have developed formulations for BNS's in 
quasi-equilibrium.  The formulation for the first and second 
post-Newtonian approximations 
(1PN/2PN, respectively) have been developed and 
co-rotating quasi-equilibrium configurations of BNS's in the 1PN 
approximation have been calculated \cite{as967}.  Also, the 
Wilson--Mathews formulation has been implemented by several authors 
\cite{wm956,ba97} to express quasi-equilibrium BNS's as 
well as space-time around them.  
In this paper we use the Wilson--Mathews formulation to express the 
gravitational field.  For the fluid, the property (B) is implemented by 
introducing the velocity potential and assuming the barotropic flow.  
The property (A) is imposed by decomposing the variables into 
the direction of the Killing vector $\vec{\ell}$ and the spatial 
direction \cite{sh98,te98,bas978}. 

Consequently the basic equations we use in this paper are identical 
with those used by Bonazzola, Gourgoulhon and Marck \cite{bgm989}
and by Marronetti, Mathews and Wilson  \cite{mmw989}.  
We will summarize the formulation and the basic equations below.  
Geometrical units with $G=c=1$ are used throughout this paper.  
Latin indices run from 1 to 3, whereas Greek
indices run from 0 to 3. 

\subsection{Formulation for the gravitational field}

We adopt the ADM decomposition for the convenience of numerical 
computations.  The metric is written as, 
\begin{eqnarray}
ds^2 &=& g_{\mu\nu} dx^\mu dx^\nu \nonumber \\[2mm]
     &=& - \alpha^2 dt^2 + 
      \gamma_{ij}\left(dx^i - \omega^i dt\right)
                 \left(dx^j - \omega^j dt\right), 
\end{eqnarray}
where $g_{\mu\nu}$, $\alpha$, $\omega^i$ and $\gamma_{ij}$ are the 4-metric, 
the lapse function, the shift vector and the 3-metric on a 
3D spatial hypersurface $\Sigma_t$, respectively.  
The unit normal to $\Sigma_t$ is written as 
\begin{equation}
n^\mu = \left( \frac{1}{\alpha}, \frac{\omega^i}{\alpha} \right) 
\quad {\rm and} \quad
n_\mu = \left( -\alpha, 0,0,0 \right), 
\end{equation}
and $\gamma_{\mu\nu}$ is defined as  
\begin{equation}
\gamma_{\mu\nu} = g_{\mu\nu} +n_\mu n_\nu.  
\end{equation}

By using the (3+1) formalism, the Einstein equations are decomposed into 
constraint equations and evolution equations.  The Hamiltonian 
constraint and the momentum constraints are, respectively, 
\begin{equation} \label{ham1}
R - K_{ij} K^{ij} + K^2  = 16 \pi \rho_H,
\end{equation}
\begin{equation} \label{mom1}
D_j K^{ij} - D^i K = 8 \pi j^{i},  
\end{equation}
where the source terms $\rho_H$ and $j^{\sigma}$ are defined by
\begin{equation} \label{rhoH}
\rho_H = n^\mu n^\nu T_{\mu\nu}, 
\end{equation}
\begin{equation} \label{j}
j^{\sigma} = - \gamma^{\sigma}_{~\mu}n_{\nu} T^{\mu\nu}. 
\end{equation}
Here $R$, $K_{ij}$, $K$, $D_i$ and $T_{\mu\nu}$ are the scalar curvature, 
the extrinsic curvature, the trace part of $K_{ij}$, the covariant 
derivative with respect to $\gamma_{ij}$, and the stress-energy tensor, 
respectively.  

According to the (3+1) formalism, the 3-metric $\gamma_{ij}$ satisfies 
the dynamical equations 
\begin{equation}
{\partial \over \partial t} \gamma_{ij} 
= - 2 \alpha K_{ij} - D_i \omega_j - D_j \omega_i.
\end{equation}
These equations are decomposed into their trace part and trace free parts 
\cite{sn95} :
\begin{equation}
{\partial \over \partial t} \ln \gamma^{1/2} = - \alpha K - D_i \omega^i,
\end{equation}
\begin{eqnarray} \label{tracefree}
\gamma^{1/3} {\partial \over \partial t} 
\left(\gamma^{-1/3} \gamma_{ij}\right) & = &
 - 2 \alpha \left(K_{ij} - \frac{1}{3} \gamma_{ij} K \right)  
\nonumber \\[2mm]
 -{} D_i \omega_j & - & D_j \omega_i + \frac{2}{3} \gamma_{ij} D_k \omega_k.
\end{eqnarray}
where $\gamma = \det \gamma_{ij}$ and $K = K^i_{~i}$.

The dynamical equations for $K_{ij}$ are also derived and decomposed 
into their trace part and trace free parts.  We only write the 
trace part :  
\begin{eqnarray} \label{traceK}
{\partial K \over \partial t} = 
\alpha R - D^iD_i\alpha + \alpha K^2 - \omega^i D_i K 
\nonumber \\[2mm]
-{} 4\pi\alpha \left(3\rho_H - S \right), 
\end{eqnarray}
where $S$ is defined as 
\begin{equation} \label{sous}
S = \gamma^{ij} T_{ij}.  
\end{equation}

The post-Newtonian approach up to the 2PN order admits exact equilibrium 
states of BNS's \cite{as967}.  However, the 2PN equations for gravity 
require a significant amount of computation.  Also, the problem of 
GR formulation for quasi-equilibrium states has not been settled.  
To make the problem tractable, we assume that the spatial part of 
the metric $\gamma_{ij}$ is conformally flat for all the time 
during the inspiraling stage as follows: 
\begin{equation} \label{spflat}
\gamma^{-1/3} \gamma_{ij} = f_{ij}, 
\end{equation}
\begin{equation} \label{confo}
\gamma_{ij} = \gamma^{1/3} f_{ij} \equiv \Psi^4 f_{ij}.
\end{equation}
where $f_{ij}$ is the flat space metric and $\Psi$ is 
a conformal factor \cite{wm956,ba97}.  
This is a rather strong restriction for the space-time since 
it is well known that the 3-metric is not conformally flat 
even for stationary axisymmetric space-time. However, 
under this assumption, we can compute exactly BNS's up to the 1PN 
order and spherically symmetric configurations in full GR.  
It is also expected that this choice minimizes the gravitational 
wave content of the (physical) space-time by removing the dynamical 
or ``wave'' degrees of freedom from the field. Moreover, since this 
choice can always be adopted to find initial data without any approximation, 
our solutions give initial conditions for 3D simulations of BNS 
coalescence in full GR \cite{su99}.  

For the time slicing condition we choose maximal slicing: 
\begin{equation} \label{maxslice}
K = 0, 
\end{equation}
and we assume that this condition holds for all the time of the evolution
\begin{equation} \label{maxslicet}
{\partial K \over \partial t} = 0, 
\end{equation}
although the other evolution equations for $K_{ij}$ are omitted 
artificially.  

From the above assumptions Eqs.~(\ref{spflat}) -- (\ref{maxslicet}), 
the basic equations for the gravitational field, become as follows.  
The scalar curvature reduces to 
\begin{equation} \label{curva}
R = - 8 \Psi^{-5} \nabla^2 \Psi, 
\end{equation}
where $\nabla^2=\nabla^i \nabla_i=f^{ij}\nabla_i\nabla_j$ 
is the Laplacian of flat 3-space.
Eqs.~(\ref{ham1}), (\ref{mom1}), (\ref{tracefree}) and 
(\ref{traceK}) are rewritten as 
\begin{equation} \label{ham2}
\nabla^2 \Psi = - \frac{1}{8} \Psi^{-7} \widetilde{K}_{ij} \widetilde{K}^{ij}
        - 2\pi\Psi^5 \rho_H ,
\end{equation}
\begin{equation} \label{mom2}
\nabla^2 \omega^i + \frac{1}{3} \nabla^i \nabla_j \omega^j
= - 2 \nabla_j \left( \alpha \Psi^{-6} \right) \widetilde{K}^{ij}
 - 16 \pi \alpha \Psi^4 j^i ,
\end{equation}
\begin{equation} \label{k2}
\widetilde{K}^{ij} = - \frac{\Psi^6}{\alpha}
   \left( \frac{1}{2}\left(\nabla^i \omega^j + \nabla^j \omega^i \right)
  - \frac{1}{3} f^{ij} \nabla_k \omega^k \right) ,
\end{equation}
\begin{equation} \label{lap2}
\nabla^2 \left(\alpha \Psi\right) = \alpha \Psi \left( \frac{7}{8} \Psi^{-8}
\widetilde{K}_{ij} \widetilde{K}^{ij}  + 2 \pi \Psi^4 (\rho + 2 S) \right) .
\end{equation}
Here we define $\widetilde{K}^{ij}$ as
\begin{equation}
\widetilde{K}^{ij} = \Psi^{10} K^{ij},
\end{equation}
and its indices are lowered by the flat metric $f_{ij}$, as 
$\widetilde{K}_{ij} = f_{ik}f_{jl}\widetilde{K}^{kl}$ .
$\nabla_i$ is the derivative with respect to the flat metric $f_{ij}$.  

To make Eq.~(\ref{mom2}) more tractable, the shift vector 
$\omega^i$ is decomposed into the sum of a vector and the gradient 
of a scalar as 
\begin{equation} \label{shf0}
\omega^i = G^i - \frac{1}{4} \nabla^i B.
\end{equation}
Eq.~(\ref{mom2}) is divided into the following equations 
for these variables $G^i$ and $B$ as, 
\begin{equation} \label{shf1}
\nabla^2 G^i
= - 2 \nabla_j \left( \alpha \Psi^{-6} \right) \widetilde{K}^{ij} 
- 16 \pi \alpha \Psi^4 j^i ,
\end{equation}
\begin{equation} \label{shf2}
\nabla^2 B = \nabla_i G^i.
\end{equation}

The basic variables of the gravitational potentials for the actual 
computation are then $\Psi$, $\alpha$, $G^i$ and $B$, and the 
corresponding 
equations are Eqs.~(\ref{ham2}), (\ref{lap2}), (\ref{shf1}) and 
(\ref{shf2}), together with relations (\ref{k2}) and (\ref{shf0}).   
It should be noted that all of these equations are elliptic type 
PDEs.  We can solve these equations iteratively with appropriate 
boundary conditions by using a certain computational method, 
which is described in a later section.  

\subsection{Formulation for the fluid part}

General relativistic irrotational flows have been considered as the  
extension of the Newtonian case and applied for accretion onto 
a black hole \cite{tcm} .  
The formulation of irrotational flow for inspiraling BNS's in 
quasi-equilibrium has been derived independently by Shibata \cite{sh98} 
and by Teukolsky \cite{te98}.  We will follow their formulation.  

The energy momentum tensor for a perfect fluid can be written as
\begin{equation}
T_{\mu\nu} = \rho\left(1 + \varepsilon + \frac{P}{\rho}\right)u_\mu u_\nu
+ P g_{\mu\nu},   
\end{equation}
where $\rho$, $\varepsilon$, $P$ and $u_\mu$ are the rest mass density, 
the specific internal energy, the pressure and the 4-velocity, 
respectively.  We assume a polytropic EOS for the fluid
\begin{equation}\label{eos}
P = \left(\Gamma - 1 \right)\rho\varepsilon = \kappa \rho^{1+1/n} ,
\end{equation}
where $n$ is the polytropic index, $\Gamma = 1 + 1/n$, and $\kappa$ is 
a constant.  We define the relativistic specific enthalpy $h$ as 
\begin{equation}
h = 1 + \varepsilon + \frac{P}{\rho} = 
1+ \kappa \left( n+1 \right)\rho^{1/n} . 
\end{equation}

The conservation equation for the energy momentum tensor 
\begin{equation} \label{econ}
\fnabla_\mu T^{\mu}_{\nu} = 0 ,
\end{equation}
can be rewritten as Euler's equation 
\begin{equation}\label{euler}
u^\mu {} \fnabla_\mu \left(h u_\nu \right) +  \fnabla_\nu h = 0 ,
\end{equation}
where $\fnabla_\mu$ is the covariant derivative with respect to 
$g_{\mu\nu}$. 
The conservation of rest mass density is written as
\begin{equation}\label{baryon}
\fnabla_\mu \left(\rho u^\mu \right) = 0.  
\end{equation}

The covariant definition of irrotationality in 4D is 
\begin{eqnarray}
\omega_{\mu\nu} & = & 
P^\alpha_{~\mu} P^\beta_{~\nu} 
\left( \fnabla_\beta u_\alpha - \fnabla_\alpha u_\beta \right) 
\nonumber \\[2mm]
& = & \frac{1}{h} 
\left( \fnabla_\nu \left(h u_\mu\right) 
- \fnabla_\mu \left(h u_\nu \right)  \right) = 0, 
\end{eqnarray}
where $P^\mu_{~\nu} = g^\mu_{~\nu} + u^\mu u_\nu$ is the projection 
tensor and we have made use of Eq.~(\ref{euler}).  
Thus the quantity $h u_\mu$ can be expressed in terms of 
the gradient of a potential $\Phi$ as 
\begin{equation}\label{vepo}
h u_\mu = \fnabla_\mu \Phi.  
\end{equation}
In the following, we refer to $\Phi$ as the velocity potential. 
Then the equation of rest mass conservation (\ref{baryon}) is 
rewritten as 
\begin{equation}\label{vepoeq}
\fnabla^\mu {} \fnabla_\mu \Phi = 
-  \fnabla^\mu \left(\ln \frac{\rho}{h} \right) \fnabla_\mu \Phi.  
\end{equation}
From the normalization condition for the 4-velocity $u_\mu u^\mu = -1$, 
we get the following equation for $h$ 
\begin{equation}\label{bernoulli}
h = \left[- \fnabla_\mu\Phi \fnabla^\mu\Phi \right]^{1/2} . 
\end{equation}
This is the generalized Bernoulli equation in GR since 
it is also derived by integrating Euler's equation (\ref{euler}) 
as follows, 
\begin{eqnarray}
\lefteqn{h \left(u^\mu {} \fnabla_\mu \left(h u_\nu \right) +
 \fnabla_\nu h\right) }
\nonumber \\[2mm]
&& =  \frac{1}{2} \fnabla_\nu 
\left( \fnabla_\mu \Phi \fnabla^\mu \Phi + h^2 \right) = 0.  
\end{eqnarray}

Now we decompose Eqs. (\ref{vepoeq}) and (\ref{bernoulli}) 
into the ADM (3+1) form and impose the 
quasi-equilibrium condition.  The quasi-equilibrium condition is 
implemented by assuming the existence of a timelike Killing 
vector $\ell^\mu$ such that 
\begin{equation}\label{quasi1}
\fnabla_\nu \ell_\mu + \fnabla_\mu \ell_\nu=0 
\end{equation}
and
\begin{equation}\label{quasi2}
{\cal L}_\ell f = \ell^\mu {} \fnabla_\mu f = 0 
\end{equation}
where $f$ denotes a certain fluid variable 
such as $\rho$, $h$ or $u_\mu$.  ${\cal L}_\ell$ denotes 
the Lie derivative along $\ell^\mu$.   Since we have assumed 
a polytropic EOS and irrotational flow, the number of independent 
fluid variables is two, i.e. the velocity potential $\Phi$ and, 
for example, the density $\rho$.  Note however that not $\Phi$ but its
gradient is the fluid variable.  Therefore, the following relation holds
\begin{equation}\label{quasi3}
{\cal L}_\ell (h u) = {\cal L}_\ell (d \Phi) = 
d({\cal L}_\ell \Phi) = 0, 
\end{equation}
that is, 
\begin{equation}\label{quasi4}
{\cal L}_\ell \Phi = \ell^\mu {} \fnabla_\mu \Phi = - C
\end{equation}
where $C$ is a positive `global' constant on the fluid.  

Performing the (3+1) decomposition of Eqs.~(\ref{vepoeq}) and 
(\ref{bernoulli}), and imposing the quasi-equilibrium conditions 
(\ref{quasi2}) and (\ref{quasi4}), we have the following basic equations 
for irrotational binary systems, respectively \cite{sh98,te98},
\begin{eqnarray}\label{vepoeq2}
D^iD_i\Phi &-& B^iD_i\frac{\lambda}{\alpha^2} - \frac{\lambda K}{\alpha}
\nonumber \\[2mm]
&=& - \left(D^i\Phi - \frac{\lambda}{\alpha^2}B^i \right)
D_i\ln\frac{\alpha\rho}{h},
\end{eqnarray}
\begin{equation}\label{bernoulli2}
h^2 = \frac{\lambda^2}{\alpha^2}- D_i\Phi D^i\Phi .  
\end{equation}
Here, $B^i$ is a rotating shift vector defined as 
\begin{equation}\label{rotshift}
\ell^\mu = \alpha n^\mu  + B^\mu, \qquad 
B^\mu = - \omega^\mu + \Omega \xi^\mu, 
\end{equation}
where $\xi^\mu$ is the generator of rotations about the rotation axis 
of the binary.  $\lambda$ is defined as 
\begin{equation}\label{lamdef}
\lambda = C + B^j D_j\Phi .
\end{equation}
Using the quasi-equilibrium conditions Eqs.~(\ref{quasi2}) and 
(\ref{quasi4}), we obtain the following expressions for the normal 
derivatives of a scalar quantity $f$ and the velocity potential $\Phi$,   
\begin{eqnarray}\label{norderi1}
n^\mu {} \fnabla_\mu f & = &
\frac{1}{\alpha}\left( \ell^\mu - B^\mu \right)\fnabla_\mu f
\nonumber \\[2mm]
&=& - \frac{1}{\alpha} B^i D_i f, 
\end{eqnarray}
and 
\begin{eqnarray}\label{norderi2}
n^\mu {} \fnabla_\mu \Phi & = &
\frac{1}{\alpha}\left( \ell^\mu - B^\mu \right)\fnabla_\mu \Phi
\nonumber \\[2mm]
 & = &
 - \frac{1}{\alpha}\left( C + B^i D_i \Phi \right)
 = -\frac{\lambda}{\alpha}. 
\end{eqnarray}

The boundary condition for the velocity potential equation 
is derived as follows from the fact that the spatial fluid velocity 
on the stellar surface flows along the stellar surface on each 
3D spatial hypersurface.  We decompose $u^\mu$ as 
\begin{equation}\label{vdeco}
u^\mu = u^0 \left(\ell^\mu + V^\mu \right) .
\end{equation}
and assume that $V^\mu$ is a spatial vector, $n_\mu V^\mu = 0$.
By using $V^\mu$, the boundary condition is expressed as 
\begin{equation}\label{bouv1}
\left. V^i D_i \rho \right|_{\rm surf} = 0  .  
\end{equation}
From the definition for $V^\mu$,  we obtain
\begin{eqnarray} \label{vv1}
 n_\mu V^\mu & = & \frac{1}{u^0} n_\mu u^\mu - n_\mu \ell^\mu 
\nonumber \\[2mm]
& = & \frac{1}{u^0} \left(- \frac{\lambda}{\alpha h} \right) + \alpha
= 0 .
\end{eqnarray}
Therefore $V^\mu$ is rewritten as follows.  
\begin{eqnarray} \label{vv2}
 V^\mu & = & \gamma^\mu_{~\nu} V^\nu 
\nonumber \\[2mm]
& = & \frac{1}{u^0} \gamma^\mu_{~\nu} u^\nu - \gamma^\mu_{~\nu} \ell^\nu 
\nonumber \\[2mm] 
& = & \frac{\alpha^2}{\lambda} \left(D^\mu \Phi - 
      \frac{\lambda}{\alpha^2} B^\mu \right) ,
\end{eqnarray}
where Eqs.~(\ref{rotshift}), (\ref{norderi2}) and (\ref{vv1}) are used.  
From this expression, the boundary condition (\ref{bouv1}) is 
written as 
\begin{equation} \label{bouvep}
\left. \left(D^i \Phi - 
\frac{\lambda}{\alpha^2} B^i \right) D_i \rho \right|_{\rm surf} = 0  .
\end{equation}
This is identical to the one derived from imposing the 
finiteness of the r.h.s. of Eq.~(\ref{vepoeq2}) on the compact support 
of a star which assures that the elliptic PDE for $\Phi$ is 
well-defined on it.  

Finally, we impose the spatial conformal flatness Eq.~(\ref{confo}) and 
the maximal slicing condition Eq.~(\ref{maxslice}).  
By using the following identities for the derivative of a scalar function,
\begin{equation}
D^i D_i \Phi = \Psi^{-4} \nabla^2 \Phi + 
2 \Psi^{-5} \nabla^i\Psi \nabla_i\Phi, 
\end{equation}
\begin{equation}
D_i A = \nabla_i A \quad {\rm and} \quad 
D^i A D_i B = \Psi^{-4} \nabla^i A \nabla_i B , 
\end{equation}
the basic equations and boundary conditions for the fluid part, namely, 
Eqs.~(\ref{vepoeq2}), (\ref{bernoulli2}),  and 
(\ref{bouvep}) are rewritten as follows, 
\begin{eqnarray}\label{vepoeq3}
\nabla^i \nabla_i \Phi = 
&-& \frac{2}{\Psi} \nabla^i \Psi \nabla_i \Phi + 
\Psi^4 B^i \nabla_i \frac{\lambda}{\alpha^2} 
\nonumber \\[2mm]
&-& \left(\nabla^i\Phi - \frac{\lambda}{\alpha^2} \Psi^4 B^i \right)
\nabla_i\ln\frac{\alpha\rho}{h},
\end{eqnarray}
\begin{equation}\label{bernoulli3}
h^2 = \frac{\lambda^2}{\alpha^2} - \Psi^{-4} \nabla_i\Phi \nabla^i\Phi ,
\end{equation}
and 
\begin{equation} \label{bouvep2}
\left. \left(\nabla^i \Phi - \frac{\lambda}{\alpha^2} 
\Psi^4 B^i \right) \nabla_i \rho \right|_{\rm surf} = 0  ,
\end{equation}
where , 
\begin{equation}\label{lamdef2}
\lambda = C + B^j \nabla_j\Phi , \qquad B^i =  \Omega \xi^i - \omega^i.   
\end{equation}
We also need another boundary condition to determine the stellar surface.
This comes from the definition of the surface itself, 
\begin{equation} \label{bousurf1}
P = 0.  
\end{equation}
Eqs.~(\ref{vepoeq3}) and (\ref{bouvep2}) form a system of 
elliptic type PDEs with Neumann boundary conditions.

Next we derive the matter source terms which appear in 
the Einstein equations, i.e. $\rho_H$, $j^\sigma$ and $S$.  
From the definitions, 
\begin{eqnarray}\label{maso1}
\rho_H &=& n^\mu n^\nu T_{\mu\nu} \,=\,
\frac{\rho\lambda^2}{h\alpha^2} - P ,
\\[2mm]\label{maso2}
S &=& \gamma^{\mu\nu}T_{\mu\nu} \,=\,
\frac{\rho \lambda^2}{h \alpha^2} - \rho h + 3P  ,
\\[2mm]\label{maso3}
j^i &=& -\gamma^{i\mu} n^\nu T_{\mu\nu} \,=\, 
\frac{\rho\lambda}{h\alpha}D^i\Phi \,=\,
\frac{\rho\lambda}{h\alpha} \Psi^{-4} \nabla^i \Phi  .
\end{eqnarray}

\section{The Computational method for irrotational binary systems in GR}

In this section, we describe the solution method for the basic 
equations derived in the previous section. This is essentially the 
extension of the method developed previously by the authors and 
co-workers and successfully applied to solving for the structure of 
a rapidly rotating single star, or a binary system in Newtonian gravity 
or in GR \cite{ue98,ha86,keh89}.  

\subsection{Coordinates and symmetries}\label{cosy}

For the present computation, we prepare two spherical coordinate systems 
$(r,\theta,\varphi)$, 
one for the gravitational field whose origin is at the intersection of 
the rotational axis and the equatorial plane, and the other for the fluid 
variables whose origin is at the coordinate center of the star.  We will 
call them the gravitational coordinates and the fluid coordinates, 
respectively.  When necessary, we will distinguish these two coordinates 
by subscripts $g$ and $f$ as $(r_g,\theta_g,\varphi_g)$ and 
$(r_f,\theta_f,\varphi_f)$, respectively.  

We assume that the symmetry of the physical quantities is as shown in 
Table \ref{tabsym}.  Note that the equatorial plane of the star is the
$\theta_g = \theta_f = \pi/2$ plane and that the axis of orbital motion 
is located at $\theta_g = 0$.  Planes of (anti-)symmetry are the 
equatorial plane and a plane with $\varphi_g = 0$ for 
the gravitational coordinates, 
$\varphi_f = 0$ and $\pi$ for the fluid coordinates.  
In this paper, we only consider an equal mass BNS system and hence 
a plane of (anti-)symmetry with respect to $\varphi_g = \pi/2$ 
exists for the gravitational coordinates as well.  

\begin{table}
\begin{center}
\begin{tabular}{ccccc}
& $\theta_g = \pi/2$ & $\varphi_g = 0$ & $\varphi_g = \pi/2$ & 
$r_g \rightarrow \infty$  \\[1mm]
\tableline
$\Psi$ & sym & sym & sym & 
$\displaystyle 1+{\cal O}\left(\frac{1}{r}\right)$ \\[3mm]
$\alpha$ & sym & sym & sym & 
$\displaystyle 1+{\cal O}\left(\frac{1}{r}\right)$ \\[3mm]
$G^r$ & sym & anti-sym & anti-sym & 
$\displaystyle {\cal O}\left(\frac{1}{r^2}\right)$ \\[3mm]
$G^\theta$ & anti-sym & anti-sym & anti-sym & 
$\displaystyle {\cal O}\left(\frac{1}{r^2}\right)$ \\[3mm]
$G^\varphi$ & sym & sym & sym & 
$\displaystyle {\cal O}\left(\frac{1}{r^2}\right)$ \\[3mm]
$\sin\varphi\, G^\varphi$ & sym & anti-sym & sym & 
$\displaystyle {\cal O}\left(\frac{1}{r^2}\right)$ \\[3mm]
$B$ & sym & anti-sym & anti-sym & 
$\displaystyle {\cal O}\left(\frac{1}{r}\right)$ \\[1mm]
\tableline
& $\theta_f = \pi/2$ & $\varphi_f = 0$ & $\varphi_f = \pi$ & 
$r_f = R_S$  \\[1mm]
\tableline
$\rho$ & sym & sym & sym & Eq.~(\ref{bousurf1}) \\[3mm]
$\Phi$ & sym & anti-sym& anti-sym & Eq.~(\ref{bouvep2}) \\[1mm]
\end{tabular}
\end{center}
\caption{Symmetries of variables and their boundary conditions 
are listed : `sym' and `anti-sym' denote plane symmetry and 
anti-symmetry, respectively.  Boundary conditions are at the outer 
boundaries, i.e. $r_g \rightarrow \infty$ for the gravitational
potentials and $r_f = R_S$ for the fluid, where $R_S$ is the stellar 
surface.  
The equatorial plane is taken to be the $\theta = \pi/2$ 
plane and the stars are aligned along the 
$(\theta_g,\varphi_g)=(\pi/2,0)$-axis.\label{tabsym}}
\end{table}

\begin{figure}
\epsfxsize=8truecm
\begin{center}
\epsffile{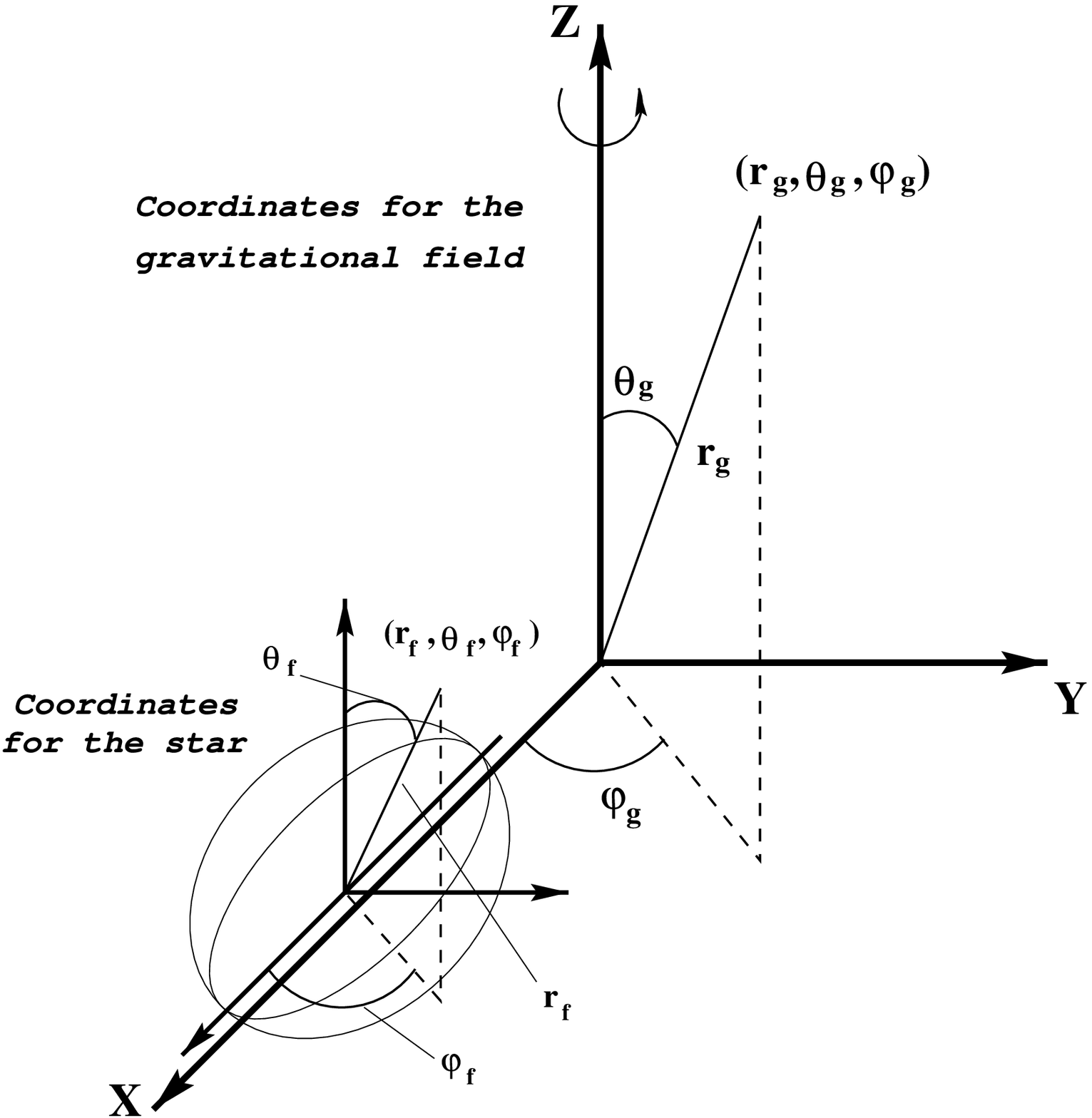}
\end{center}
\epsfxsize=8truecm
\begin{center}
\epsffile{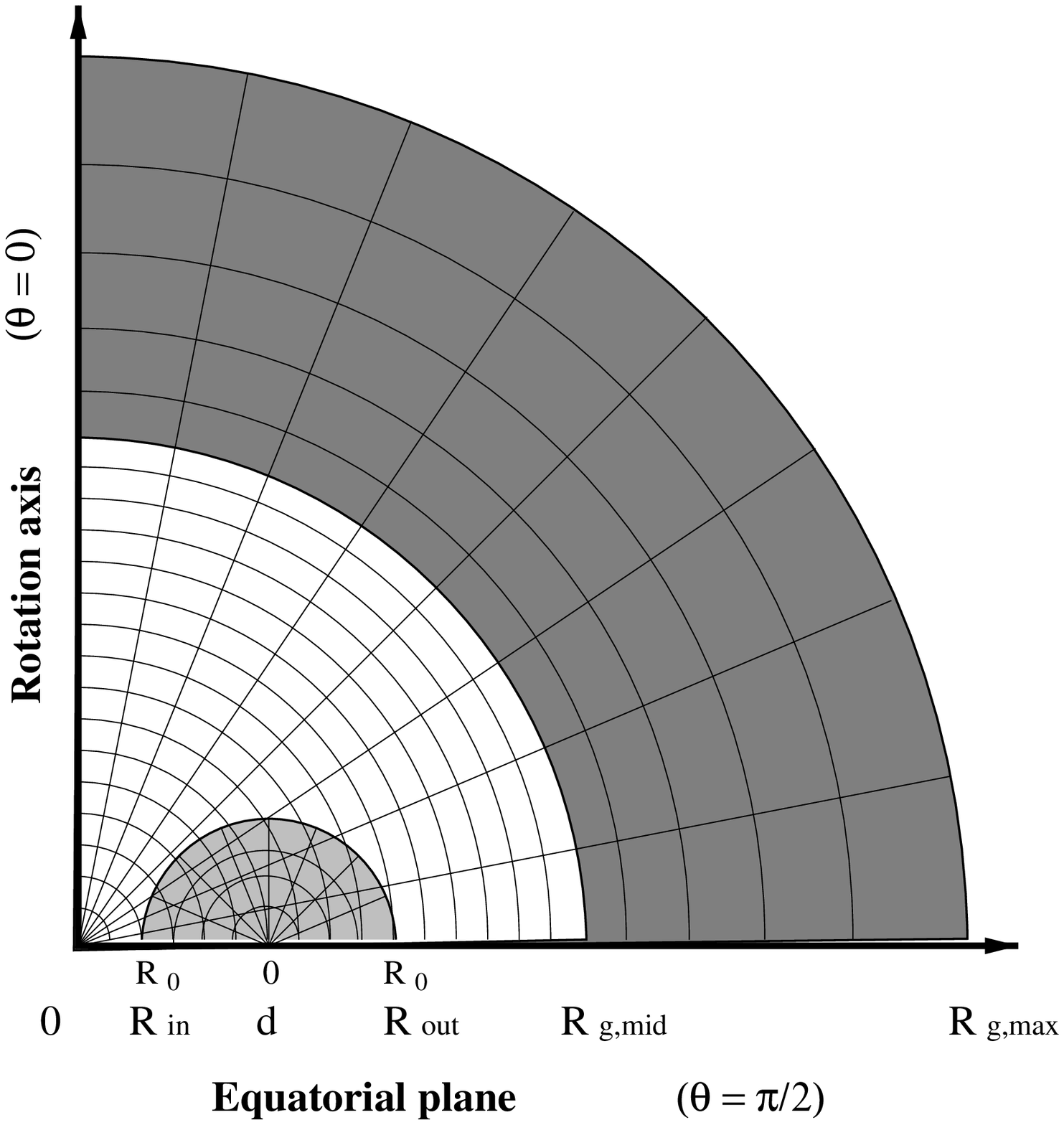}
\end{center}
\caption{Schematic figure of the coordinate systems used in the actual 
computation.  
(a) A sketch of the coordinate systems for the star and 
the gravitational field. 
(b) The larger spherical coordinate system corresponds to that for 
gravitation (white and dark gray region).  
The smaller spherical coordinate system corresponds to that for the
fluid (gray region).  $R_0$ is defined as $R_0 = (R_\ouou - R_\inin)/2$.  
\label{figponzu}}
\end{figure}

Because of these symmetries, we only need to treat one octant 
of the spherical coordinate system for the gravitation, namely,
$(r_g,\theta_g,\varphi_g)\in [0,\infty)\times[0,\pi/2]\times[0,\pi/2]$.  
For the fluid coordinate system, we need to treat a quarter of 
the whole star, that is, 
$(r_f,\theta_f,\varphi_f)\in [0,R]\times[0,\pi/2]\times[0,\pi]$, 
where $R$ is the largest coordinate radius of the star in the fluid 
spherical coordinates.  We set the coordinate center of the star to be 
at $(r_g,\theta_g,\varphi_g)=(d,\pi/2,0)$ of the gravitational 
coordinate system, where 
\begin{equation} \label{sepa}
d = (R_\ouou + R_\inin)/2 ,
\end{equation}
is defined as the coordinate center of the star, 
and $(r_g,\theta_g,\varphi_g)=(R_\ouou,\pi/2,0)$ and
$(R_\inin,\pi/2,0)$ are the outer 
and inner edges of the star, respectively.  We show a schematic 
figure of the coordinate systems in Figure \ref{figponzu}.  

Since we have assumed conformal flatness in space, we may introduce 
these spherical coordinate systems in the same manner as we do for 
Newtonian models.  Especially, since the derivatives appearing in 
the basic equations are those for the flat 3-metric $f_{ij}$, 
it is convenient to take the non-coordinate basis whose metric 
becomes the unit matrix $f_{ij} = 
\delta_{ij} = {\rm diag}(1,1,1)$ as is usually done for vector analysis 
in flat space, where $\delta_{ij}$ is the Kronecker delta.  
We write the orthonormal non-coordinate basis for 
the spherical coordinates as 
\[
\{{\bf e}_r, {\bf e}_\theta, {\bf e}_\varphi \}, \quad {\rm where}
\]
\begin{equation} 
{\bf e}_a \cdot {\bf e}_b = \delta_{ab} , \quad (a,b = r,\theta,\varphi) .
\end{equation} 
When necessary, we will distinguish the bases for the gravitational 
coordinate system and the fluid coordinate system by the superscripts 
$g$ and $f$, 
respectively.  Using these bases, the gradient of a scalar function 
$\phi$ is written as 
\begin{equation} \label{nabnab}
\nabla^i \phi = \nabla_i \phi = 
{\bf e}_r \frac{\partial \phi}{\partial r} + 
{\bf e}_\theta \frac{1}{r} \frac{\partial \phi}{\partial \theta} + 
{\bf e}_\varphi\frac{1}{r\sin\theta} \frac{\partial \phi}{\partial \varphi}.  
\end{equation}
Hereafter, we often for convenience make use of the index notation and 
the notation of vector analysis in the same equation, as above.  

\subsection{Poisson solver}\label{secproc}

Seven elliptic type PDEs appear in our formulation. 
They are Eqs.~(\ref{ham2}), (\ref{lap2}), (\ref{shf1}) and (\ref{shf2}) 
for the gravitational field, and Eq.~(\ref{vepoeq3}) for the fluid 
variable.  To solve these equations, the flat part of 3D Laplacian 
is first separated out.  Then, the equations are transformed into 
integral forms by using the Green's function of the 3D Laplacian.   
\begin{equation} \label{exeli}
\Delta \phi_{(a)} = S_{(a)} . 
\end{equation}
Here the suffix $(a)$ is an index to specify each equation, and 
$\Delta$ is the 3D flat Laplacian in spherical coordinates, 
\begin{eqnarray} \label{sphlap}
\lefteqn{
\Delta \phi_{(a)} = 
\frac{1}{r^2}\frac{\partial}{\partial r}
\left(r^2\frac{\partial \phi_{(a)}}{\partial r}\right) +} \nonumber \\[2mm]
&&\frac{1}{r \sin\theta}\frac{\partial}{\partial \theta}
\left(\sin\theta \frac{\partial \phi_{(a)}}{\partial \theta} \right) + 
\frac{1}{r^2 \sin^2 \theta}\frac{\partial^2 \phi_{(a)}}{\partial \varphi^2}. 
\end{eqnarray}
By using a Green's function which satisfies the following equation
\begin{equation}
\Delta\left(-\frac{1}{4\pi}
\frac{1}{\left|\bf{r}-\bf{r'}\right|}\right) = 
\delta(\bf{r}-\bf{r'}), 
\end{equation}
Eq.~(\ref{exeli}) is transformed as 
\begin{equation} \label{exint}
\phi_{(a)} = -\frac{1}{4\pi} \int \frac{S_{(a)}}
{\left|\bf{r}-\bf{r'}\right|} dV' + \chi, 
\end{equation}
where $\chi$ is either a function or a constant depending on 
the boundary condition.  
We use the Legendre expansion to compute the integral 
in Eq.~(\ref{exint}), 
%
\begin{eqnarray} \label{legen}
\lefteqn{
\frac{1}{\left|\bf{r}-\bf{r'}\right|}\,=\, 
\sum_{n=0}^\infty f_n(r,r') \sum_{m=0}^n \epsilon_m \,
\frac{(n-m)!}{(n+m)!}\,
} \nonumber\\[2mm]
&&
\times P_n^{~m}(\cos\theta)\,P_n^{~m}(\cos\theta')
\cos m(\varphi-\varphi'), 
\end{eqnarray}
%
where $f_n(r,r')$ is defined as,
\begin{equation}
f_n(r,r')  =
\left\{
\begin{array}{cc}
\displaystyle 
\frac{1}{r}\left(\frac{r'}{r}\right)^n, 
\quad {\rm for} \quad r' \le r \ ,  \\ \\
\displaystyle 
\frac{1}{r'}\left(\frac{r}{r'}\right)^n, 
\quad {\rm for} \quad r \le r' \ , 
\end{array}
\right.
\end{equation}
$\epsilon_m$ is defined as,
\begin{equation}
\epsilon_m = 
\left\{
\begin{array}{ll}
 1 \ ,\quad {\rm for} \quad m\,=\,0\ ,  \\ \\
 2 \ ,\quad {\rm for} \quad m\,=\,1,2,...,n, 
\end{array}
\right.
\end{equation}
and $P_n^{~m}(\cos\theta)$ is the associated Legendre function.  
Here $dV' = r'^2 \sin\theta' dr'd\theta' d\varphi'$.  

Eq.~(\ref{exint}) is a formal solution in a sense that each 
source term $S_{(a)}$ also contains the variables $\{\phi_{(a)}\}$.  
Therefore, an iterative method is used to find a solution.  
All of the equations are consistently solved by a self-consistent-field 
(SCF) iteration scheme \cite{ha86,keh89,om68}, whose procedure is 
described in detail in Paper I of \cite{ue98}.  

\subsection{Solution method for the gravitational field}

To apply the above procedure, we need to separate out the 3D flat 
Laplacian $\Delta$ for the spherical coordinates as in Eq.~(\ref{sphlap}).  
We may regard $\nabla^2$ operating on scalar valued functions, 
as in Eqs.~(\ref{ham2}), (\ref{lap2}) and (\ref{shf2}), 
in the same way as $\Delta$ in Eq.~(\ref{sphlap}).  
On the other hand, the Laplacian $\nabla^2$ operates on a vector 
valued function on the l.h.s.~of Eq.~(\ref{shf1}).  
Although it is possible to use the Green's function of the Laplacian 
for a vector valued function, instead, we explicitly write down its 
components and derive the form of Eq.~(\ref{sphlap}) in each expression, 
so that we can follow the same procedure as for the scalars.  

In the gravitational coordinates, this procedure goes as follows,
\begin{equation}\label{lapg0}
\nabla^2 G^j = \nabla^i \nabla_i G^j = 
\Delta \left( G^r {\bf e}_r^g + G^\theta {\bf e}_\theta^g + 
G^\varphi {\bf e}_\varphi^g \right), 
\end{equation}
\begin{eqnarray}
\label{lapgr}
&&\left[\nabla^2 G^j \right]^r = 
{\bf e}_r \cdot \Delta \left( G^r {\bf e}_r^g + 
G^\theta {\bf e}_\theta^g + G^\varphi {\bf e}_\varphi^g \right)
\nonumber \\[2mm]
&& = \Delta G^r - \frac{2}{r^2}\left( 
  G^r
+\frac{G^\theta}{\tan\theta} 
+\frac{1}{\sin\theta}\frac{\partial G^\varphi}{\partial \varphi}
+\frac{\partial G^\theta}{\partial \theta} \right), 
\\[4mm]
\label{lapgt}
&&\left[\nabla^2 G^j\right]^\theta = 
{\bf e}_\theta \cdot \Delta \left( G^r {\bf e}_r^g + 
G^\theta {\bf e}_\theta^g + G^\varphi {\bf e}_\varphi^g \right)
\nonumber \\[2mm]
&& = \Delta G^\theta -  \frac{2}{r^2}\left( 
 \frac{G^\theta}{2 \sin^2\theta}
+\frac{\cos\theta}{\sin^2\theta}
\frac{\partial G^\varphi}{\partial \varphi}
-\frac{\partial G^r}{\partial \theta} \right), 
\\[4mm]
\label{lapgp}
&&\left[\nabla^2 G^j\right]^\varphi = 
{\bf e}_\varphi \cdot \Delta \left( G^r {\bf e}_r^g + 
G^\theta {\bf e}_\theta^g + G^\varphi {\bf e}_\varphi^g \right)
\nonumber \\[2mm]
&& = \Delta G^\varphi - \frac{2}{r^2}\left( 
 \frac{G^\varphi}{2 \sin^2\theta}
-\frac{1}{\sin\theta}\frac{\partial G^r}{\partial \varphi}
-\frac{\cos\theta}{\sin^2\theta}
\frac{\partial G^\theta}{\partial \varphi} \right), 
\end{eqnarray}
where Eqs.~(\ref{lapgr}) $\sim$ (\ref{lapgp}) are the l.h.s.~of 
Eq.~(\ref{shf1}).  
For the $\varphi$ component of Eq.~(\ref{shf1}), 
we multiply by $\sin\varphi$ and rearrange the terms of 
Eq.~(\ref{lapgp}) as follows: 
\begin{eqnarray}\label{lapgpp}
\sin\varphi\left[\nabla^2 G^j\right]^\varphi &=& 
\sin\varphi \,{\bf e}_\varphi \cdot \Delta \left( G^r {\bf e}_r^g + 
G^\theta {\bf e}_\theta^g + G^\varphi {\bf e}_\varphi^g \right) 
\nonumber \\[2mm]
=\Delta(\sin\varphi\, G^\varphi) &-& \frac{2}{r^2 \sin^2\theta}\left( 
-\sin\theta\sin\varphi\frac{\partial G^r}{\partial \varphi} \right.
\nonumber \\[2mm]
&-&\left.
\cos\theta\sin\varphi\frac{\partial G^\theta}{\partial \varphi}
+\cos\varphi\frac{\partial G^\varphi}{\partial \varphi} \right). 
\end{eqnarray}
In these expressions we suppress the subscript $g$ for the gravitational
coordinates.  

Terms appearing in expressions (\ref{lapgr}), (\ref{lapgt}) 
and (\ref{lapgpp}) in addition to the flat 3D Laplacian $\Delta$ are 
transferred to the r.h.s.~and are included as a part of the source 
term of each 
component.  By the above rearrangement, all of the basic equations for 
the gravitational part have been written in the form of Eq.~(\ref{exeli}).  
Therefore, we can follow the procedure described in subsection 
\ref{secproc}.  As we show in Table \ref{tabsym}, the boundary conditions 
are imposed by setting $\chi=1$ for Eqs.~(\ref{ham2}) and 
(\ref{lap2}) and $\chi=0$ for Eqs.~(\ref{shf1}) and (\ref{shf2}).  
Because of the symmetries for the variables shown in Table \ref{tabsym}, 
the source terms have corresponding symmetries as well.  As a result, 
appropriate behaviors of integral terms in Eq.~(\ref{exint}) 
at $r \rightarrow \infty$ are automatically satisfied.  

\subsection{Solving method for the fluid part}

The basic equations and boundary conditions for the fluid part are 
Eqs.~(\ref{vepoeq3}) -- (\ref{bousurf1}).  Since these equations 
are essentially the same as those of the Newtonian case solved in 
Paper I, we can apply all of the computational techniques developed 
in that paper.  As we mentioned in subsection \ref{cosy}, we prepare 
the spherical coordinate 
system to compute the structure of the star whose origin is at the 
coordinate center of the star.  The elliptic type PDE 
Eq.~(\ref{vepoeq3}) is solved in this coordinate system by using 
the same procedure as in subsection \ref{secproc}.  Eq.~(\ref{vepoeq3}) 
has the same form as Eq.~(\ref{exeli}) and is integrated to give 
Eq.~(\ref{exint}).  
To impose the boundary condition Eq.~(\ref{bouvep2}), we regard the 
term $\chi$ in Eq.~(\ref{exint}) as a regular homogeneous solution of 
the Laplace equation inside the star, i.e., 
\begin{equation}\label{homoge}
\Delta \chi(r_f,\theta_f,\varphi_f) = 0.  
\end{equation}
In the spherical coordinates, such a homogeneous solution 
to Eq.~(\ref{homoge}) can be expressed by using the associated 
Legendre functions as follows:
\begin{eqnarray}\label{sph}
&& \chi(r_f,\theta_f,\varphi_f)\,=\,\sum_{l=1}^\infty\sum_{m=1}^l\,
a_{lm}\,r_f^l\,\left[1+(-1)^{l+m}\right]\, \nonumber \\[2mm]
&&\times \left(\frac{(2l+1)(l-m)!}{4\pi(l+m)!}\right)^{1/2}\,
P_l^{~m}(\cos\theta_f)\,\sin m\varphi_f\ ,
\end{eqnarray}
where $a_{lm}$'s are certain constants.  
Here we have taken into account the regularity of the solutions
at $r_f=0$ and the symmetries listed in Table \ref{tabsym}.  
The coefficients $a_{lm}$ are computed so that the velocity potential 
$\Phi$ satisfies the boundary condition (\ref{bouvep2}).  

Eq.~(\ref{bernoulli3}) is used to compute the density distribution 
within the star.  For numerical computations, we introduce a function 
$q$ which is similar to the Emden function of Newtonian polytropes 
defined as follows:
\begin{equation}\label{emden}
q = \frac{P}{\rho} .
\end{equation}
In terms of this function, we can express 
\begin{eqnarray}\label{emden2}
\rho &=& \kappa^{-n} q^n ,  \\[2mm]
P &=& \kappa^{-n} q^{n+1} , \\[2mm]
h &=& 1 + \left( n+1 \right) q .
\end{eqnarray}
Substituting the last three expressions into Eq.~(\ref{bernoulli3}), 
we find 
\begin{equation}\label{bernoulli4}
q = \frac{1}{n+1}\left[ \left( \frac{\lambda^2}{\alpha^2} - 
\Psi^{-4} \nabla_i\Phi \nabla^i\Phi \right)^{1/2} - 1 \right] .
\end{equation}

Accordingly, the matter source terms Eqs.~(\ref{maso1})--(\ref{maso3})
are rewritten as follows,
\begin{eqnarray}\label{masokp1}
\rho_H &=& \kappa^{-n} q^n 
\left(\frac{\lambda^2}{\left( 1+\left(n+1 \right)q \right) \alpha^2}
 - q \right),
\\[2mm]\label{masokp2}
\rho_H &+& 2 S \,=\, \kappa^{-n} q^n 
\nonumber \\[2mm]
&\times&  \left(
\frac{3\lambda^2}{\left(1+\left(n+1\right)q \right)\alpha^2}
- \left(2 + \left(2n-3 \right)q\right)
\right),
\\[2mm]\label{masokp3}
j^i &=& \kappa^{-n} q^n 
\frac{\lambda}{\left( 1+\left(n+1 \right)q \right) \alpha}
\Psi^{-4} \nabla^i \Phi.
\end{eqnarray}

For the numerical computation of the fluid part, it is convenient 
to introduce the surface fitted spherical coordinate system 
as used in the Newtonian case \cite{ue98}.  
The surface of one NS in the binary system can be 
expressed by a function $R_S(\theta_f,\varphi_f)$ even when
the deformation of the shape is relatively large.  
By using this function, the new coordinates $(r_f^*,\theta_f^*,\varphi_f^*)$ 
are defined as follows, 
\begin{equation}\label{sfco}
r_f^* = \frac{r_f}{R_S(\theta_f,\varphi_f)}, \quad
\theta_f^* = \theta_f, \quad {\rm and } \quad
\varphi_f^* = \varphi_f. 
\end{equation}
The stellar interior with the assumed symmetry is mapped into the 
region 
$(r_f^*,\theta_f^*,\varphi_f^*)\in [0,1]\times[0,\pi/2]\times[0,\pi]$.  
The surface fitted coordinate is advantageous for computing 
numerical derivatives and imposing boundary conditions
at the stellar surface.  
Implementation of this coordinate system is described in detail 
in Paper I. 
\subsection{Normalization of quantities and choice of parameters}

Non-dimensionalization of variables and proper choices of parameters 
are important for the iteration scheme to obtain converged 
equilibrium configurations stably.  
For convenience in making numerical computations, we rescale the 
coordinate length so that the coordinate radius of the star at the 
intersection of the surface and the coordinate line 
with $(\theta_g,\varphi_g)=(\pi/2,0)$ is unity, namely, 
we introduce a parameter $R_0$ so that 
$R_S(\pi/2,0)/R_0 = 1 = R_S(\pi/2,\pi)/R_0$ in the fluid coordinates.  

By using this $R_0$, the basic equations for the gravitational 
potentials, Eqs.~(\ref{ham2}), (\ref{lap2}), (\ref{shf1}) and 
(\ref{shf2}), whose forms are essentially the same as 
Eq.~(\ref{exeli}) or Eq.~(\ref{exint}), become as follows, 
\begin{equation} \label{exeli2}
\widehat{\Delta} \phi_{(a)} = 
\widehat{S}^{~g}_{(a)}  + \kappa^{-n} R_0^{~2} \, \widehat{S}^{~m}_{(a)} , 
\end{equation}
\begin{equation} \label{exint2}
\phi_{(a)} = \chi
 - \frac{1}{4\pi} \int \frac{\widehat{S}_{(a)}^{~g} + 
\kappa^{-n}R_0^{~2}\,\widehat{S}_{(a)}^{~m}}
{\left|\widehat{\bf{r}}\,-\,\widehat{\bf{r}}'\right|} d\widehat{V}' ,
\end{equation}
where $\widehat{\Delta} = R_0^{~2}\,\Delta$.  Here $\widehat{S}^{~m}_{(a)}$ 
stands for the non-dimensionalized part of the matter source terms 
corresponding to Eqs.~(\ref{maso1})--(\ref{maso3}) or 
(\ref{masokp1})--(\ref{masokp3}) and 
$\widehat{S}^{~g}_{(a)}$ denotes remaining terms.  We explicitly 
show the dependence of these quantities on the parameters $\kappa$ 
and $R_0$.  In the equations for the fluid part, $\kappa$ and $R_0$ 
are canceled out after rescaling and non-dimensionalization.  
By this rescaling, the coordinates and variables are transformed as 
follows, 
\begin{eqnarray}\label{paraP}
\widehat{r} = \frac{r}{R_0}, \quad
\widehat{\nabla}_i = R_0 \nabla_i, \quad
\widehat{\xi}^i = \frac{\xi^i}{ R_0}, \quad
\widehat{R}_S = \frac{R_S}{R_0},
\nonumber \\[2mm]
\widehat{K}_{ij} = \widetilde{K}_{ij} R_0, \quad
\widehat{B} = \frac{B}{R_0}, \quad
\widehat{\Phi} = \frac{\Phi}{R_0}, \quad
\widehat{\Omega} = \Omega R_0.
\end{eqnarray}

We may consider that $\kappa$ is used for non - dimensionalization of 
the variables, and $R_0$ for rescaling of the numerical computations.  
However, since these two parameters appear in the equations only through 
the following combination:
\begin{equation} \label{paraR}
\bar{R}_0 = \kappa^{-n/2} R_0 , 
\end{equation}
we can regard them as a single parameter when we compute each solution.  
In this paper, we only consider $\kappa = {\rm const}$ sequences, which 
could be appropriate for investigation of the final inspiraling stage 
of BNS's.  (When constructing solutions, we could mimic 
physical effects such as heating by changing the value of $\kappa$.)  

We then have three parameters $\bar{R}_0$, $\widehat{\Omega}$ and 
$C$ in the basic equations.  We need to impose three more conditions to 
specify them.  For this purpose, we choose three locations where $q$ has 
definite values.  Namely, we set $q=0$ at 
the intersections of the surface and the coordinate line 
with $(\theta_g,\varphi_g)=(\pi/2,0)$, i.e. at the two points 
$(r_f^*,\theta_f^*,\varphi_f^*) = (1,\pi/2,0)$ and $(1,\pi/2,\pi)$ 
and also set $q = q_c$ at a grid point where $q$ takes the largest 
value inside the star \cite{qccomment} .
Since we have introduced the surface 
$\widehat{R}_S(\theta_f^*,\varphi_f^*)$ in the surface fitted 
coordinates, 
the former two conditions are explicitly imposed by setting 
$\widehat{R}_S(\pi/2,0) = 1 = \widehat{R}_S(\pi/2,\pi)$.  

The above three conditions are applied to Eq.~(\ref{bernoulli3}) 
at three points to get equations for $\bar{R}_0$, $\widehat{\Omega}$ and $C$.  
It is known \cite{keh89,ba97} that $\alpha$ and $\Psi$ 
are scaled as 
\begin{equation}\label{scale2}
\alpha = \widehat{\alpha}^{\bar{R}_0^2}, \qquad
\Psi = \widehat{\Psi}^{-\bar{R}_0^2/2}.
\end{equation}
By using this rescaling, Eq.~(\ref{bernoulli3}) is written as, 
\begin{eqnarray}\label{bernoulli5}
\widehat{\alpha}^{2 \bar{R}_0^2}\left\{1+\left(n+1\right) q \right\}^2
+ \left(\widehat{\alpha}\widehat{\Psi}\right)^{2 \bar{R}_0^2}
\widehat{\nabla}_i\widehat{\Phi}\,
\widehat{\nabla}^i\widehat{\Phi}
\nonumber \\[2mm] 
- \left\{C + \left(\widehat{\Omega}\,\widehat{\xi}^i - \omega^i \right)
\widehat{\nabla}_i\widehat{\Phi} \right\}^2 = 0.   
\end{eqnarray}
We use three equations which are derived by imposing 
the above three conditions 
at three points in the star, namely, at the inner and outer edges
and at the coordinate center of the star.  
These three non-linear algebraic equations with respect to 
$\bar{R}_0$, $\widehat{\Omega}$ and $C$ are solved by using 
the Newton-Raphson method and their values are updated through 
the iteration procedure.  

The above choice is known to make the iteration converge stably.  
This choice of parameters and the computation scheme for them are 
essentially the same 
as those for rotating single stars \cite{keh89} or those for co-rotating 
BNS's \cite{ba97} in GR. 

\subsection{Discretization and numerical computation}

Since we have rescaled the length by $R_0$, the computational domain is 
measured by this unit.  For the gravitational coordinate system, 
the whole computational domain is taken as 
$(\widehat{r}_g,\theta_g,\varphi_g)
\in[0,\widehat{R}_{g,\rm max}]\times[0,\pi/2]\times[0,\pi/2]$.  
It is discretized equidistantly for the $\theta_g$ and $\varphi_g$ 
directions.  For the $\widehat{r}_g$ direction, we divide 
the region into two parts as $[0,\widehat{R}_{g,\rm mid}]$ 
(the white region in Figure \ref{figponzu}(b)) and 
$[\widehat{R}_{g,\rm mid},\widehat{R}_{g,\rm max}]$ 
(the dark gray region in Figure \ref{figponzu}(b)).  
We discretize the region $[0,\widehat{R}_{g,\rm mid}]$ equidistantly,  
and the region $[\widehat{R}_{g,\rm mid},\widehat{R}_{g,\rm max}]$ 
non-equidistantly.  
By denoting grid points as $(\widehat{r}_{i},\theta_{j},\varphi_{k})$, 
in which we suppress the subscript $g$, the discretization becomes 
as follows,
\begin{eqnarray}
\widehat{r}_{i+1}-\widehat{r}_{i} &=& 
\frac{\widehat{R}_{g,\rm mid}}{N^r_{g,\rm mid}},
\quad {\rm where} \quad 
\nonumber \\[2mm]
0 &\le& \widehat{r}_{j} \le \widehat{R}_{g,\rm mid}, \ 
\left(0\le j \le {N^r_{g,\rm mid}}\right),
\\[2mm]
\widehat{r}_{i+1}-\widehat{r}_{i} &=& 
\delta(\widehat{r}_{i}-\widehat{r}_{i-1}),
\quad {\rm where} \quad 
\nonumber \\[2mm]
\widehat{R}_{g,\rm mid} &\le& \widehat{r}_{j} \le \widehat{R}_{g,\rm max}, 
\  
\left({N^r_{g,\rm mid}} \le j \le {N^r_{g,\rm max}}\right) ,
\label{coout}
\\[2mm]
\theta_{i+1}-\theta_{i} &=& \frac{\pi/2}{N^\theta_{g}},
\quad {\rm where} \quad 
\nonumber \\[2mm]
0 &\le& \theta_{j} \le \frac{\pi}{2}, \ 
\left(0\le j \le {N^\theta_{g}}\right),
\\[2mm]
\varphi_{i+1}-\varphi_{i} &=& \frac{\pi/2}{N^\varphi_{g}},
\quad {\rm where} \quad 
\nonumber \\[2mm]
0 &\le& \varphi_{j} \le \frac{\pi}{2}, \ 
\left(0\le j \le {N^\varphi_{g}}\right).  
\end{eqnarray}
In Eq.~(\ref{coout}), $\delta (>1)$ is a certain constant.  

The surface fitted fluid coordinate system is discretized 
equidistantly in all directions as follows, 
\begin{eqnarray}
r^*_{i+1}-r^*_{i} &=& 
\frac{1}{N^r_{f}}, 
\quad {\rm where} \quad 
\nonumber \\[2mm]
0 &\le& \widehat{r}^*_j \le 1, \ 
\left(0\le j \le {N^r_{f}}\right)
\\[2mm]
\theta^*_{i+1}-\theta^*_{i} &=& \frac{\pi/2}{N^\theta_{f}},
\quad {\rm where} \quad 
\nonumber \\[2mm]
0 &\le& \theta^*_{j} \le \frac{\pi}{2}, \ 
\left(0\le j \le {N^\theta_{f}}\right),
\\[2mm]
\varphi^*_{i+1}-\varphi^*_{i} &=& \frac{\pi}{N^\varphi_{f}},
\quad {\rm where} \quad 
\nonumber \\[2mm]
0 &\le& \varphi^*_{j} \le \pi, \ 
\left(0\le j \le {N^\varphi_{f}}\right).  
\end{eqnarray}
The fluid coordinate system is always set within the region 
$[0,R_{g,\rm mid}]$ of the gravitational coordinate system in order 
to maintain an accuracy.  
Standard finite differences are applied for the basic 
equations with these grid points, which maintain at least second 
order accuracy. 

In the present computation, we typically took the values listed in 
Table \ref{tabgrid}.  A finer mesh is used for the region 
$[0,\widehat{R}_{g,\rm mid}]$ whose size in the $\widehat{r}_g$ 
direction is about 2.5 times the diameter of one NS for 
the cases with types S, M, and L and 2 times 
for the case with type Ls. 
The region $[\widehat{R}_{g,\rm mid},\widehat{R}_{g,\rm max}]$
is discretized into $20$ grid points in the $\widehat{r}_g$ direction.  
Types S, M, and L are mainly used to show the convergence of the scheme 
as functions of the mesh size and we will mention this in a later 
section.  The mesh sizes of type S and type M in the region 
$0 \le \widehat{r_g} \le \widehat{R}_{g,\rm mid}$ are, 
respectively, $1/2$ and $3/4$ of that of the finest type L.  
Type Ls is used for compute the configurations for 
rather small separations, although the numerical results of 
type L and type Ls are almost the same. 
\begin{table}
\begin{center}
\begin{tabular}{cccccccccc}
$\widehat{R}_{g,\rm mid}$ & $\widehat{R}_{g,\rm max}$ 
& $ N^r_{g,\rm mid} $ & $ N^r_{g,\rm max} $ & $ N^\theta_{g} $ 
& $ N^\varphi_{g} $ & $ N^r_{f} $ & $ N^\theta_{f} $ 
& $ N^\varphi_{f} $ & type \\[1mm]
\tableline
$  5 $ & $ 100 $ & $ 40 $ & $ 20 $ & $ 20 $ & $ 30 $ & 
$  8 $ & $   8 $ & $ 16 $ &  S  \\[1mm]
$  5 $ & $ 100 $ & $ 60 $ & $ 20 $ & $ 30 $ & $ 45 $ & 
$ 12 $ & $  12 $ & $ 24 $ &  M  \\[1mm]
$  5 $ & $ 100 $ & $ 80 $ & $ 20 $ & $ 40 $ & $ 60 $ & 
$ 16 $ & $  16 $ & $ 32 $ &  L  \\[1mm]
$  4 $ & $ 100 $ & $ 80 $ & $ 20 $ & $ 40 $ & $ 40 $ & 
$ 20 $ & $  40 $ & $ 28 $ &  Ls  \\[1mm]
\end{tabular}
\end{center}
\caption{Parameters for the computational domains and the numbers 
of grid points in each coordinate system are listed.  
\label{tabgrid}}
\end{table}

Strictly speaking, $\widehat{R}_{g,\rm max}$ should be infinite in 
order to impose asymptotically flatness as a boundary condition.  
This could be implemented by using a certain appropriate coordinate 
transformation to compactify an infinite region, which has been 
used by several authors \cite{cst92,bgm98}.  Although this is 
straightforward, we do not implement such a special treatment for 
simplification in the present numerical code.  
Instead, we truncate the computational domain at large 
$\widehat{R}_{g,\rm max}$. It is known that such truncation does not 
affect the local properties of the space-time or the structure of 
each NS \cite{nsge98}. Although the choice of 
the value of $R_{g,max}$ would affect the values of the physical 
quantities very little, it would be safer to cover the space as 
extensively as possible.  We take a rather large value for 
$\widehat{R}_{g,\rm max} = 100$ as shown in Table \ref{tabgrid}.  
Because of this choice, the error introduced into an integral 
quantity such as the ADM mass by this truncation 
(see Eq.~(\ref{M}) below) does not affect the digits of 
the numerical results presented in the later sections. 

We also truncate the order of the Legendre expansion in 
Eq.~(\ref{legen}) and Eq.~(\ref{sph}) at finite values instead of 
infinity. We denote the maximum value 
of the expansion for quantities in the gravitational coordinates 
in Eq.~(\ref{legen}) as $n_{\rm max}$ and that for the fluid 
coordinates in Eq.~(\ref{legen}) and Eq.~(\ref{sph}) 
(i.e. the expansion for the velocity potential) as $l_{\rm max}$.  
Typically, we use $(n_{\rm max}, l_{\rm max}) = (32,12)$.
Therefore sources of truncation error are the finite difference and 
the truncation for this expansion.  The results depend 
very little on $l_{\rm max}$ when we choose $l_{\rm max} > 8$. 
We will show how the results depend on $n_{\rm max}$ 
in a later section.

As was mentioned before, the basic equations are solved iteratively.  
The iteration procedure is the same as that for Newtonian 
irrotational BNS models described in Paper I.  The only difference in 
the numerical scheme is that an interpolation of physical quantities 
from one coordinate system to the other is required.  We implemented 
the cubic spline interpolation method for all directions of the 
coordinates.  As we will show in the later sections, one of the 
interesting problems relating to the irrotational binary sequence 
is to determine the behavior of the maximum density as 
the binary stars approach.  To investigate this problem, it is quite 
important to compute a smooth density distribution.  
In fact, an interpolation scheme of 1st order accuracy does not give 
a smooth density distribution, especially around the central region.
Therefore, we have employed the cubic spline interpolation which
keeps smoothness of interpolated data, although the cost for numerical 
computations is rather large.

In our formulation, each quasi-equilibrium configuration is 
specified by three parameters, namely, the half of the separation 
between the coordinate centers of the two component stars $\widehat{d}$, 
the largest value of $q_c$ on the discretized grid, and 
the polytropic index $n$.  
(Note that $\kappa$ has been scaled out.)  
The initial guess for the numerical iteration may be, for instance, 
an analytic solution of a Newtonian spherical polytrope 
with $n=1$ at larger separation, say $\widehat{d}=2$.  We have 
started the iteration by setting $q_c$ to be a small value and could 
then follow iteration cycles.  We have regarded the obtained 
solution as being 
a converged one when relative differences of physical variables of 
subsequent iterations on each coordinate grid point are less than 
a certain small 
value, typically we choose $10^{-5}$.  Once we get a converged 
solution, we can change the values of parameters by a few tenths 
of a percent.  
Typically, a solution is converged after 100 $\sim$ 300 iteration cycles.  
It takes about $1$ minute for one iteration cycle using a DEC 
Alpha-station 1/166.  

\section{Numerical results for irrotational binary systems in GR}

\subsection{Integrals}

In this section we show several integral quantities of  
the equilibrium configurations.  The total rest-mass energy 
$M_{0,\rm tot}$ of the two component stars is defined as 
\begin{equation}
M_{0,\rm tot} = \int_{\cal M} \rho u^{\mu} d \Sigma_{\mu}
        = \int_{\cal M} \rho u^\mu \left(-n_\mu\sqrt{\gamma}\right) dV,
\end{equation}
where ${\cal M}$ denotes the integration region which is 
the support of the two component stars.  
The non-dimensional form of the total rest-mass can be written
using Eq.~(\ref{paraP}) and (\ref{paraR}) as 
\begin{eqnarray} \label{M_0}
\bar M_{0,\rm tot} = \kappa^{-n/2} M_{0,\rm tot}
&=& \kappa^{-n/2} {R}_0^3 
\int_{\cal M} \kappa^{-n} q^n \frac{\lambda}{h \alpha}\Psi^6 d\widehat{V}
\nonumber \\[2mm]
&=& \bar{R}_0^3 
\int_{\cal M} \frac{q^n \lambda \Psi^6}{h \alpha} d \widehat{V}.
\end{eqnarray}
where $d \widehat{V} = dV / R_0^3$.
The total mass-energy (ADM mass) is 
\begin{eqnarray}
M_{\rm ADM}& = &- \frac{1}{2 \pi} \oint_{\infty} \nabla^i \Psi d^2 S_i
        =  - \frac{1}{2 \pi} \int_{\infty} \nabla^2 \Psi d V,
        \nonumber \\[2mm]
& = & \frac{1}{16 \pi} \int_{\infty} \Psi^{-7}
        \widetilde K_{ij} \widetilde K^{ij} d V
        + \int_{\cal M} \Psi^5 \rho_H \,d V
\end{eqnarray}
where Eq.~(\ref{ham2}) derived from the Hamiltonian constraint is used.  
This can be rewritten in the non-dimensional form as,
\begin{eqnarray} \label{M}
\bar M_{\rm ADM} &=& \kappa^{-n/2} M_{\rm ADM} =
   \frac{\bar R_0}{16 \pi}
   \int_{\infty} \Psi^{-7} \widehat{K}_{ij} \widehat{K}^{ij} d \widehat{V}
   \nonumber \\[2mm]
   & & + \bar R_0^3 \int_{\cal M}
   \Psi^5 q^n \left( \frac{\lambda^2}{h \alpha^2} -
   q \right) d \widehat{V}.
\end{eqnarray}

The angular momentum is aligned with the $\theta_g=0$ axis and
can be defined as (see, e.g.,~\cite{by80}) 
\begin{eqnarray}
J_{\rm tot} 
&=& \frac{1}{8 \pi} \oint_{\infty}
     f_{ij} \xi^i \widetilde K^{jk} d^2 S_k
= \frac{1}{8 \pi} \int_{\infty} f_{ij} \xi^i
  \nabla_k \widetilde K^{jk} d V
\nonumber \\[2mm]
&=& \int_{\cal M} \Psi^{10} f_{ij} \xi^i j^j d V
= \int_{\cal M} \Psi^{10} r_g \sin\theta_g j^\varphi\, d V 
, 
\end{eqnarray}
where we used $\nabla_k \widetilde K^{jk} = \Psi^{10} D_k K^{jk}$ 
as well as the momentum constraints~(\ref{mom1}),   
and $\xi = r_g\sin\theta_g {\bf e}_\varphi^g$ from the definition.  
This is the total angular momentum contained
in the space-time including both the orbital and spin angular
momentum of the stars.  
Finally, we can substitute Eq.~(\ref{masokp3}) for $j^i$ and write 
the angular momentum in the non-dimensional form
\begin{eqnarray} \label{J}
\bar J_{\rm tot} & = & \kappa^{-n} J_{\rm tot} = 
\bar R_0^4 \int_{\cal M} \Psi^{10} \widehat{r}_g \sin\theta_g 
\widehat{j}^\varphi\, d \widehat{V} 
\nonumber \\[2mm]
&=& \bar R_0^4 \int_{\cal M} \Psi^{6} \widehat r_g \sin\theta_g 
\frac{q^n \lambda}{h\alpha}\widehat\nabla^\varphi\Phi\,d \widehat V .
\end{eqnarray}

In the following we will denote a half of the total rest-mass, the ADM 
mass and the angular momentum by $\bar M_0 = \bar M_{0,\rm tot}/2$, 
$\bar M = \bar M_{\rm ADM}/2$ and $\bar J = \bar J_{\rm tot}/2$,
respectively.  In the limit of 
large separation, $\bar M_0$ and $\bar M$ approach the corresponding
values for isolated stars.  

We also use the averaged separation $d_G$ which coincides with the 
separation between mass centers of the two component stars 
in Newtonian limit, namely, 
\begin{eqnarray} \label{d_G}
\bar d_{G} &=& \kappa^{-n/2} d_{G}
\nonumber \\[2mm]
&=& \frac{\bar R_0}{\bar M_0}
\int_{\cal M} \widehat r_g \sin\theta_g \cos\varphi_g 
\frac{q^n \lambda \Psi^6}{h \alpha} d \widehat V.
\end{eqnarray}

\subsection{Test of our new computational method}

In this paper we will show results for $n=1$ polytropes.
For this we need to specify two parameters in order to obtain 
one quasi-equilibrium configuration, i.e. the half of separation 
$\widehat d$ and the largest value of $q = q_c$.  
In order to check how our new computational code works, we have 
compared our results with (a) those of co-rotating GR binary systems 
tabulated in \cite{ba97}, and (b) those of the Newtonian irrotational 
binary systems in Paper I.  With the former comparison (a), we can 
check the gravitational field part of our code for the strong gravity 
regime.  We have also performed (c) a convergence test 
for the Newtonian case since it is desirable to show how the scheme 
actually converges to some analytic or semi-analytic results.  

Concerning irrotational binary systems, Taniguchi \cite{ta99} has 
analytically solved semi-detached irrotational binary systems with 
the 1PN approximation for $n=0$ polytropes and Taniguchi and Nakamura 
recently developed a method to treat the compressible Newtonian case
\cite{tn99}. Lombardi, Rasio \& Shapiro have calculated the 
irrotational polytropic binary systems by partly including 1PN 
correction terms \cite{lrs97}.  It will be important to perform the 
convergence test with these results as well as making a comparison 
with numerical results obtained by other authors who have used the 
identical formulation but different numerical computational methods.

It is convenient to characterize solutions or sequences by using the 
total rest mass $M_0 = {\rm const}$ or, equivalently, the compactness 
$(M/R)_\infty$ of an isolated spherical star. In actual computations, 
we choose a certain value for $M_0$ and adjust the parameter $q_c$ 
during the iteration so that the value of $M_0$ is kept constant.

\subsubsection{Comparison with other works}

Below, we show the results of comparison tests (a) and (b).  
In this section, type Ls in Table \ref{tabgrid} is used
for the numerical grids.  For the expansions in Eq.~(\ref{legen})
and Eq.~(\ref{sph}), we include terms up to 
$(n_{\rm max},l_{\rm max}) = (20,12)$.  

In Table \ref{tabcba}, we compare our results for co-rotating binary
systems  with those in \cite{ba97}.  For these models we have replaced 
the equations for the fluid part by those for co-rotating fluids.  
Therefore, the code for the gravitational part has been checked.  
We set the values of $\widehat{d}$ and $q_c$ to be the same as 
those in  \cite{ba97} for each model.  As shown in Table \ref{tabcba}, 
the calculated quantities $\bar \Omega$ and 
$\bar R_0$ which characterize the models for BNS's are in good 
agreement (relative differences are less than $1\%$).  
For the integral quantities $\bar M_0$, $\bar M$ and $\bar J$, 
the relative differences are at most $2\%$.  

\begin{table}
\begin{center}
\begin{tabular}{cccccccc}
$z_A$ & $\widehat d$  & $q_c$ & $\bar \Omega$ & $\bar R_0$ & 
$\bar M_0$ & $\bar M$ & $\bar J$ \\[1mm]
\tableline
0.20 & 1.50 & 0.0685 & 0.079 & 1.004 & 0.1118 & 0.10552 & 0.02729 \\[1mm]
     &      &        & 0.079 & 1.005 & 0.1125 & 0.10619 & 0.02780 \\[1mm]
\tableline
0.00 & 1.00 & 0.0658 & 0.101 & 1.177 & 0.1118 & 0.10551 & 0.02715 \\[1mm]
     &      &        & 0.101 & 1.176 & 0.1128 & 0.10637 & 0.02760 \\[1mm]
\tableline
0.20 & 1.50 & 0.2450 & 0.168 & 0.646 & 0.1781 & 0.16013 & 0.04948 \\[1mm]
     &      &        & 0.168 & 0.643 & 0.1801 & 0.16169 & 0.05050 \\[1mm]
\tableline
0.00 & 1.00 & 0.2164 & 0.202 & 0.780 & 0.1781 & 0.16018 & 0.05024 \\[1mm]
     &      &        & 0.203 & 0.774 & 0.1807 & 0.16227 & 0.05132 
\end{tabular}
\end{center}
\caption{Comparison between the results of 
Baumgarte et~al.~(1997) and those obtained with the present method 
for co-rotating GR binary systems.  The first line of each model 
corresponds to the results of Baumgarte et~al.~(1997) and 
the second line to the present results.  The first two models 
correspond to the system with $(M/R)_\infty=0.1$ and the latter 
two models to $(M/R)_\infty=0.2$ in Baumgarte et~al.~(1997).  
$z_A = 0.0$ corresponds to a contact binary system and 
$z_A = 0.2$ to a semi-detached binary system where 
$z_A = R_\inin/R_\ouou$. 
\label{tabcba}}
\end{table}

\begin{figure}
\epsfxsize=8truecm
\begin{center}
\epsffile{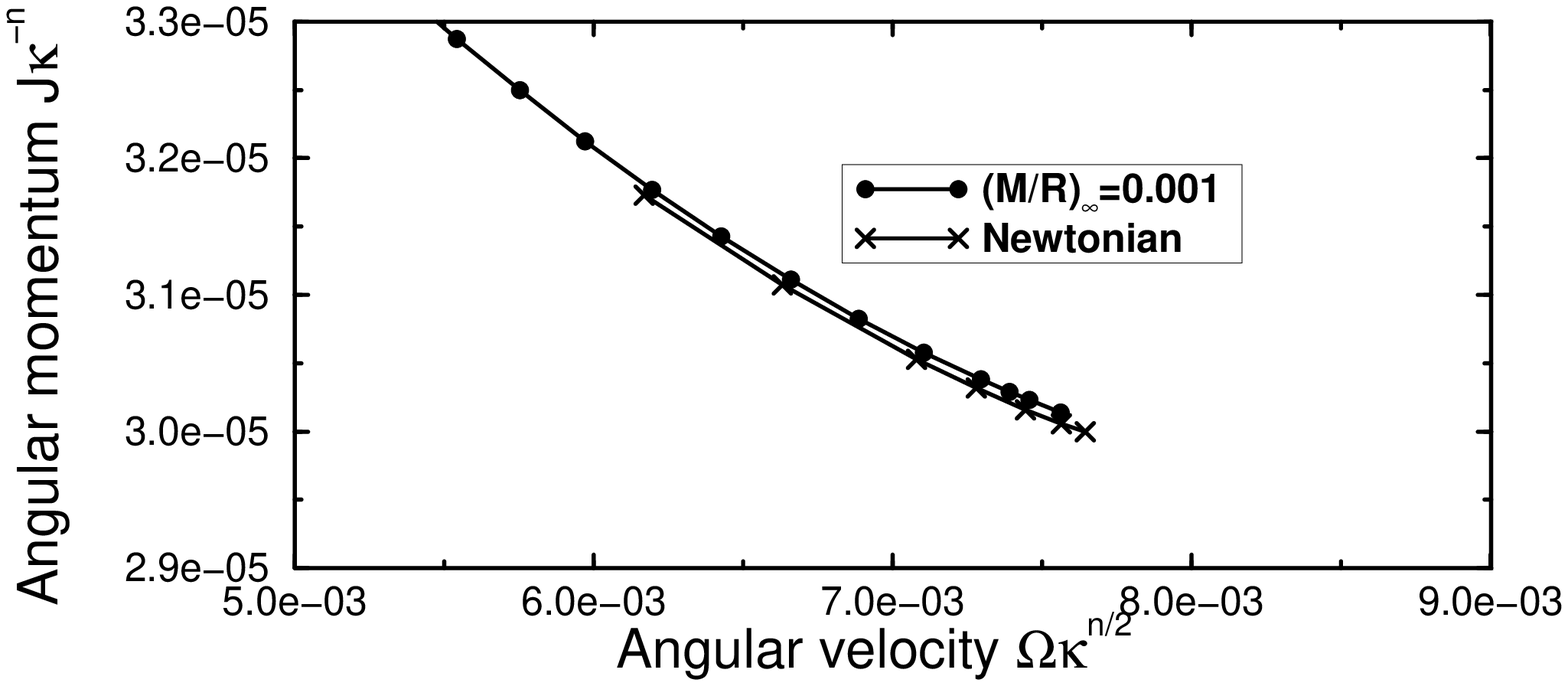}
\end{center}
\epsfxsize=8truecm
\begin{center}
\epsffile{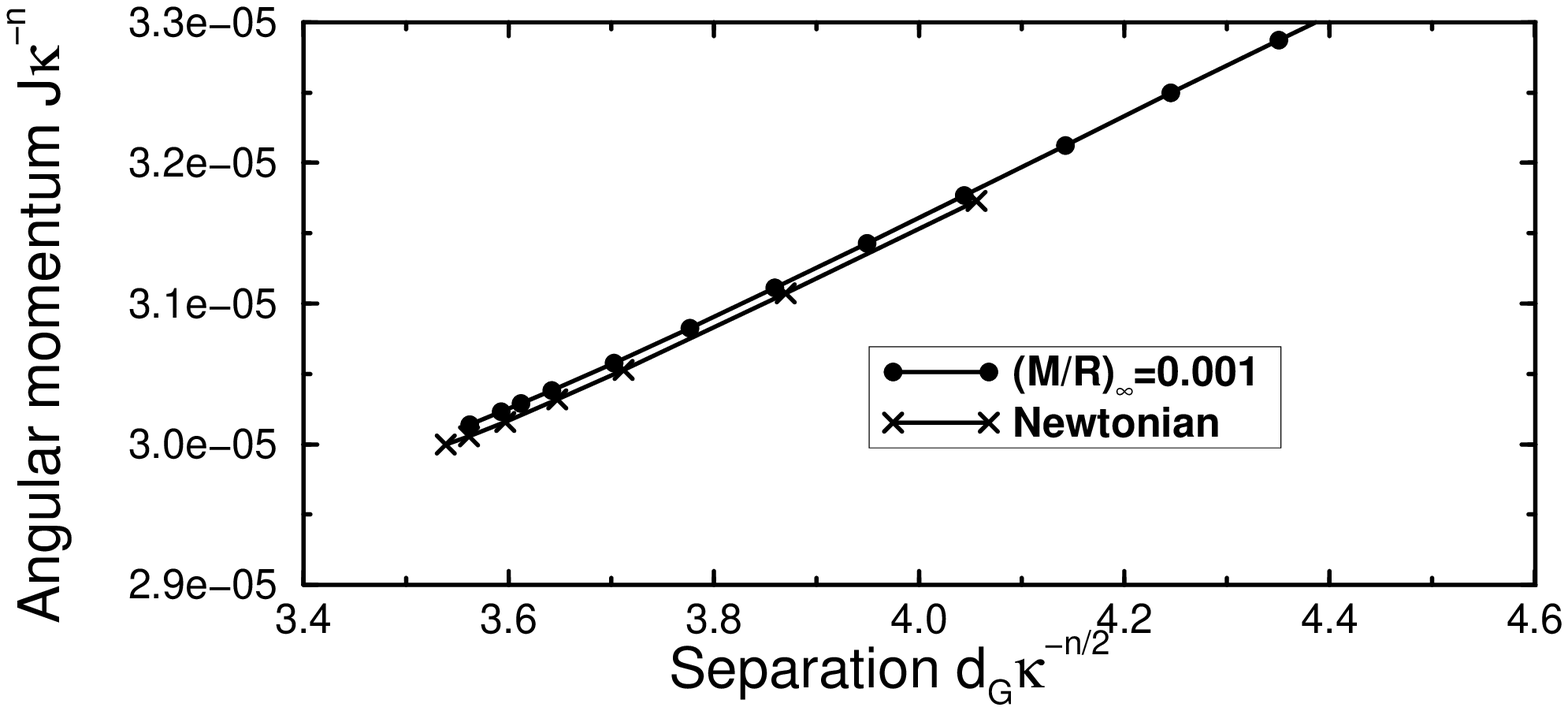}
\end{center}
\caption{Comparison between the results of Paper I (Newtonian) and 
those of the present computation (with compactness $(M/R)_\infty=0.001$).  
Quasi-equilibrium sequences for irrotational BNS systems with weak 
gravity (curve with dots) and with Newtonian gravity (curve with
crosses) are plotted.  The angular momentum as a function of the 
angular velocity (upper panel) and the angular momentum as a function 
of the separation (lower panel) are shown.  \label{fignew1}}
\end{figure}

Next, we compare our results for a quasi-equilibrium sequence of 
irrotational binary systems in weak gravity with those of Newtonian 
computations.  
It is reasonable to assume that the total rest mass is constant 
throughout the final evolution. Thus the quasi-equilibrium sequence 
with $M_0 = {\rm const}$ mimics the realistic evolution sequence as 
a result of GW emission.  
In Figure \ref{fignew1}, we compare our present results for 
the compactness $(M/R)_\infty = 0.001 $ with those in 
Paper I.  In this figure, $\bar{J}$ is plotted against the separation 
$\bar{d}_G$ or $\bar{\Omega}$.  
Relative differences of $\bar{J}$ at a fixed value of $\bar{d}_G$ or 
$\bar{\Omega}$ are $\sim 0.5\%$, relative differences of $\bar{d}_G$ at 
a fixed value of $\bar{J}$ are $\sim 1\%$, and relative differences of 
$\bar{\Omega}$ at a fixed value of $\bar{J}$ are $\sim 1.7\%$.
Therefore, our new code has reproduced the same results as those
of other computations with reasonable accuracy.  
%

\subsubsection{Convergence test}

\begin{figure}
\epsfxsize=8truecm
\begin{center}
\epsffile{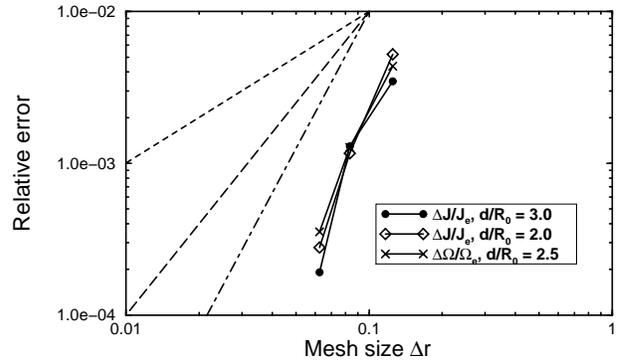}
\end{center}
\caption{Results of the convergence test for the weak gravity 
configurations (compactness $(M/R)_\infty=0.001$).  Relative 
differences between the present results and those of the semi-analytic 
calculations with the ellipsoidal approximation by Lai, Rasio and 
Shapiro (1994) are plotted. 
The points on each line correspond to the results for 
the mesh types S, M and L in Table \ref{tabgrid}. 
We plot the mesh size of the r-direction on the horizontal axis. 
The short dashed line is proportional 
to $\Delta r$, the long dashed line to $(\Delta r)^2$, and
the dash dotted line to $(\Delta r)^3$. 
\label{figconvJO}}
\end{figure}

In order to see how well our numerical solutions approximate true 
solutions, we have estimated the influence of the mesh size on the 
results.  This can be done by investigating how the values of physical 
quantities behave when the mesh size is varied. For this convergence 
test, we have used types S, M, and L in Table \ref{tabgrid} for 
numerical computations.  We chose 
$(n_{\rm max},l_{\rm max}) = (20,12)$ for the expansions.  In 
Fig.~\ref{figconvJO}, we plot the relative differences of the angular 
momentum $J$ and the angular velocity $\Omega$ from those of the 
semi-analytic results of the ellipsoidal approximation \cite{lrs94} 
for types S, M and L with values of the separation $d/R_0$ being 
fixed.  The plots clearly show that for global quantities, 
our results tend to converge as the mesh size is decreased.  
Note that even for $d/R_0=2.0$, the difference between the
polar radius and the equatorial radius is $\sim 3\%$ and this means 
that the results of the ellipsoidal approximation are reasonably 
accurate. 

On the other hand we need to be more careful about the convergence 
of local quantities such as the density distribution, the shape 
of the star, or the metric potential distribution.  Since we 
will discuss dependence of the maximum density on the binary 
separation in a later section, we here discuss the property of 
the convergence of the maximum density as a representative of 
local quantities.  

\begin{figure}
\epsfxsize=8truecm
\begin{center}
\epsffile{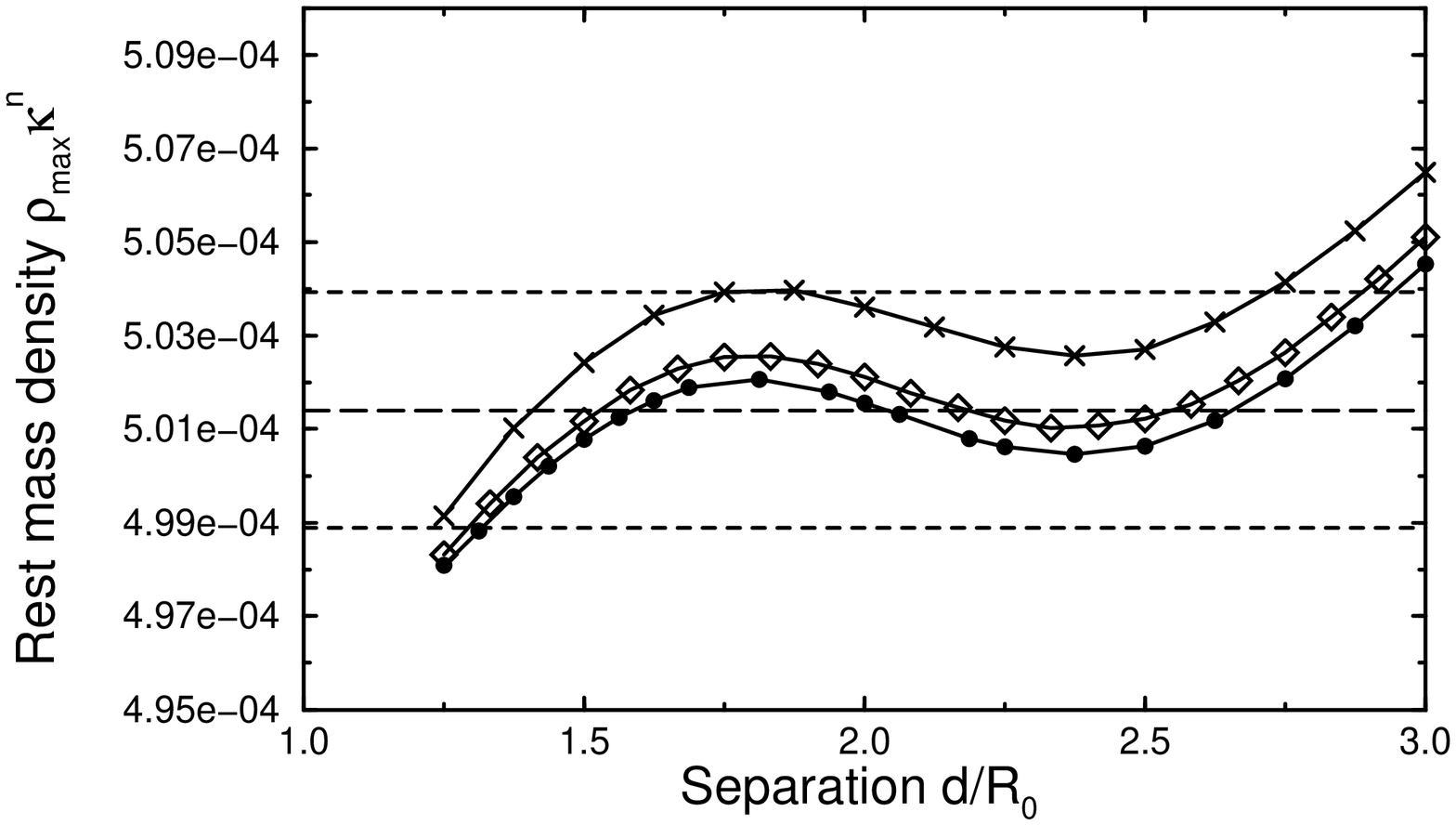}
\end{center}
\epsfxsize=8truecm
\begin{center}
\epsffile{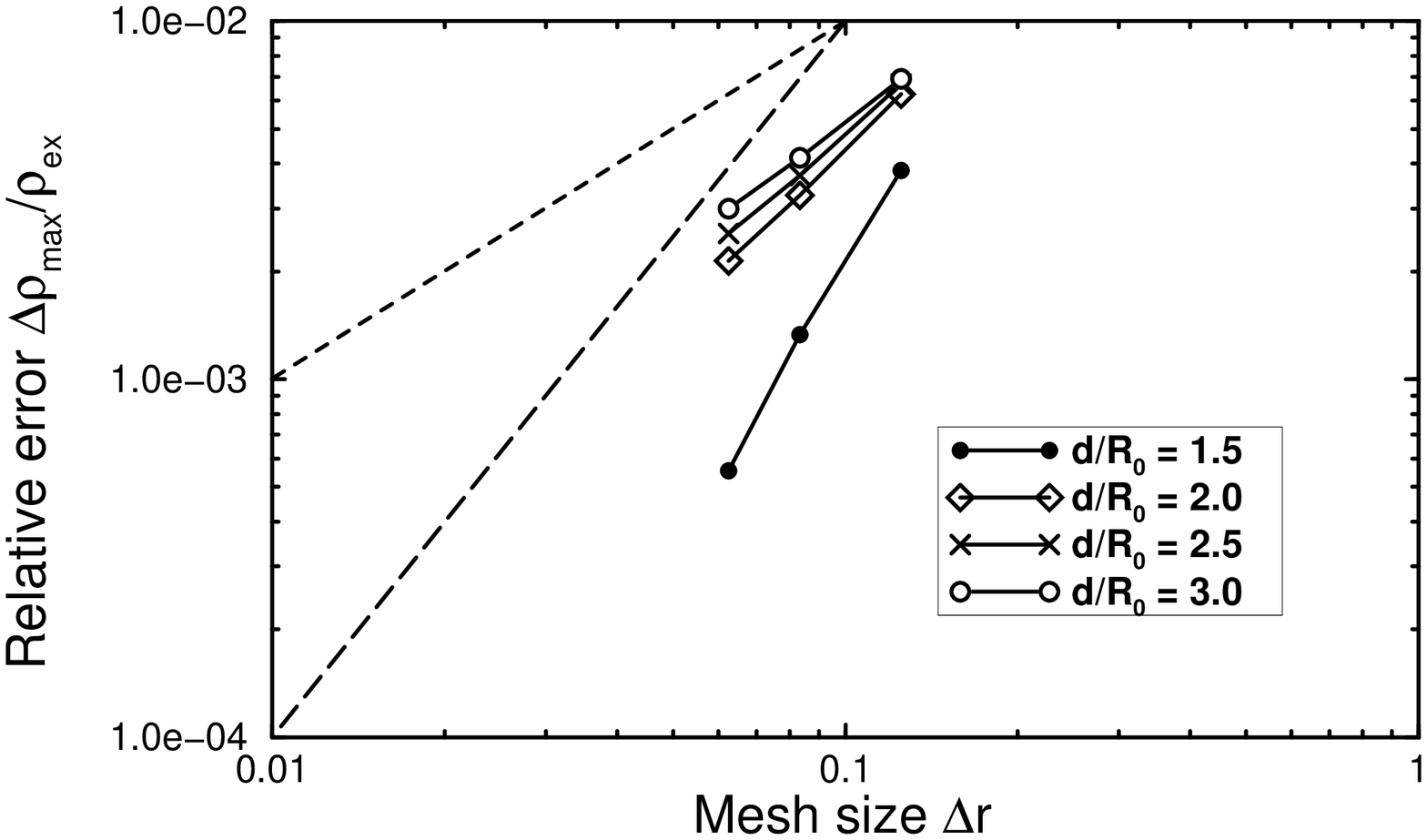}
\end{center}
\caption{The convergence test for the maximum rest mass density 
$\rho_{\rm max}$ for the case with compactness $(M/R)_\infty=0.001$.  
(a) $\rho_{\rm max}(d/R_0)$ is plotted. 
Each line corresponds to a different mesh size, 
the mesh type S (the line with crosses), 
the mesh type M (the line with diamonds), and 
the mesh type L (the line with filled circles), 
respectively. 
The long dashed line shows $\rho_{\rm max}$ of an isolated 
spherical star $\rho_{\rm max,\infty}$.  The short dashed 
lines show $\pm 0.5 \%$ difference from this value. 
(b) The convergence of the relative error for $\rho_{\rm max}$.  
The mesh size of the r-direction is used for the 
horizontal axis to represent the resolution. 
The short dashed line is proportional 
to $\Delta r$ and the long dashed line to $(\Delta r)^2$. 
\label{figconvrho}}
\end{figure}

In Fig.~\ref{figconvrho}(a), we plot the maximum rest mass density 
$\rho_{\rm max}$ against the half separation $d/R_0$, normalized by 
the $R_0$ of each model, to show the convergence of 
solutions as the mesh type is changed.  In the figure, 
the curves correspond to the mesh type S 
(the line with crosses), the mesh type M 
(the line with diamonds), and the mesh type L 
(the line with filled circles), respectively.  
We again set $(n_{\rm max},l_{\rm max}) = (20,12)$.  
The maximum density $\rho_{\rm max}(d/R_0)$ does not change 
monotonically as the separation changes, and does not tend 
exactly to the asymptotic 
value $\rho_{\rm max,\infty}$, which is the central density of an 
isolated spherical star computed from the TOV equation.  However, the 
curves oscillate around $\rho_{\rm max,\infty}$ and the differences from 
it are at most $1 \%$ even for the smallest mesh type S.  
Note that the resolution of the star in the gravity coordinates 
becomes coarser for larger separations. 
For instance, the mesh number of the gravity coordinates which 
cover the neutron star radius is only $6$ points in the $\varphi_g$ 
coordinate at $d/R_0 \sim 3.0$ for the mesh type S.  

To show that $\rho_{\rm max}(d/R_0)$ converges to a certain number for 
each separation, we plot the relative difference of $\rho_{\rm max}$, 
$\Delta \rho_{\rm max}/\rho_{\rm ex}$ against the mesh size, 
in Fig.~\ref{figconvrho}(b). 
Here the relative difference is measured in terms of the value 
$\rho_{\rm ex}$ extrapolated from the values obtained for the mesh 
types S, M and L in Table \ref{tabgrid} for each $d/R_0$. 
This figure shows that the local quantities such as the density 
also converge as the number of mesh points is increased. 

The relative difference of the maximum density is roughly 
proportional to $(\Delta r)^2$ around $d/R_0\sim 1.5$.  
This is satisfactory since we 
are interested in configurations with $d/R_0 < 2.0$,  
since the deformation of a star is small even at $d/R_0 = 2.0$.  
For example the difference between the major and minor radii is 
only a few $\%$.  Therefore, analytic or semi-analytic 
approximations could be applied to configurations down to 
this separation \cite{ta99,tn99,lrs97}.  
The results also indicate that the truncation error 
of the expansion $n_{\rm max}$ is negligible for the case with 
$d/R_0 \lo 1.5$ when we choose $n_{\rm max} \go 20$. 

\begin{figure}
\epsfxsize=8truecm
\begin{center}
\epsffile{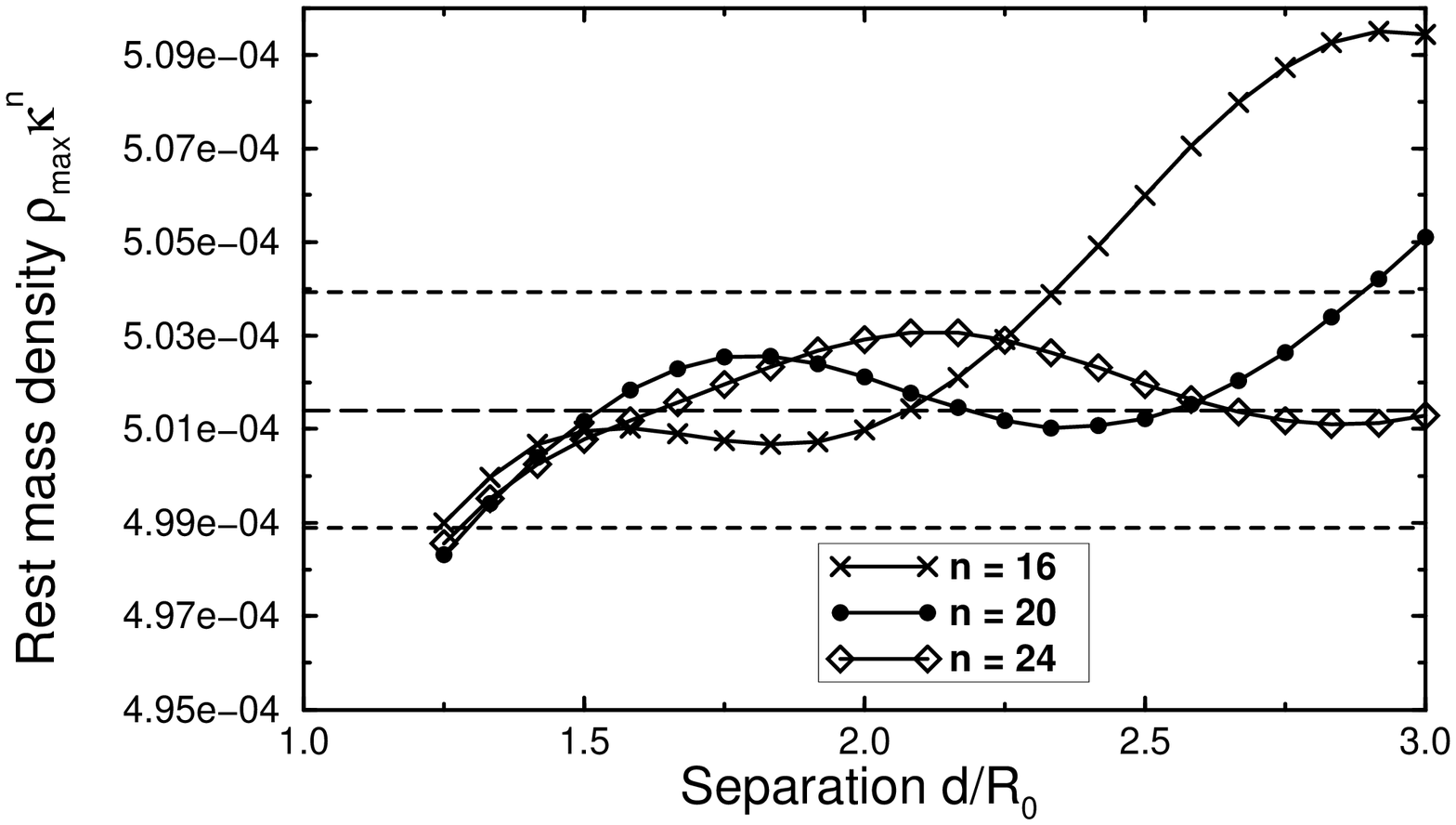}
\end{center}
\epsfxsize=8truecm
\begin{center}
\epsffile{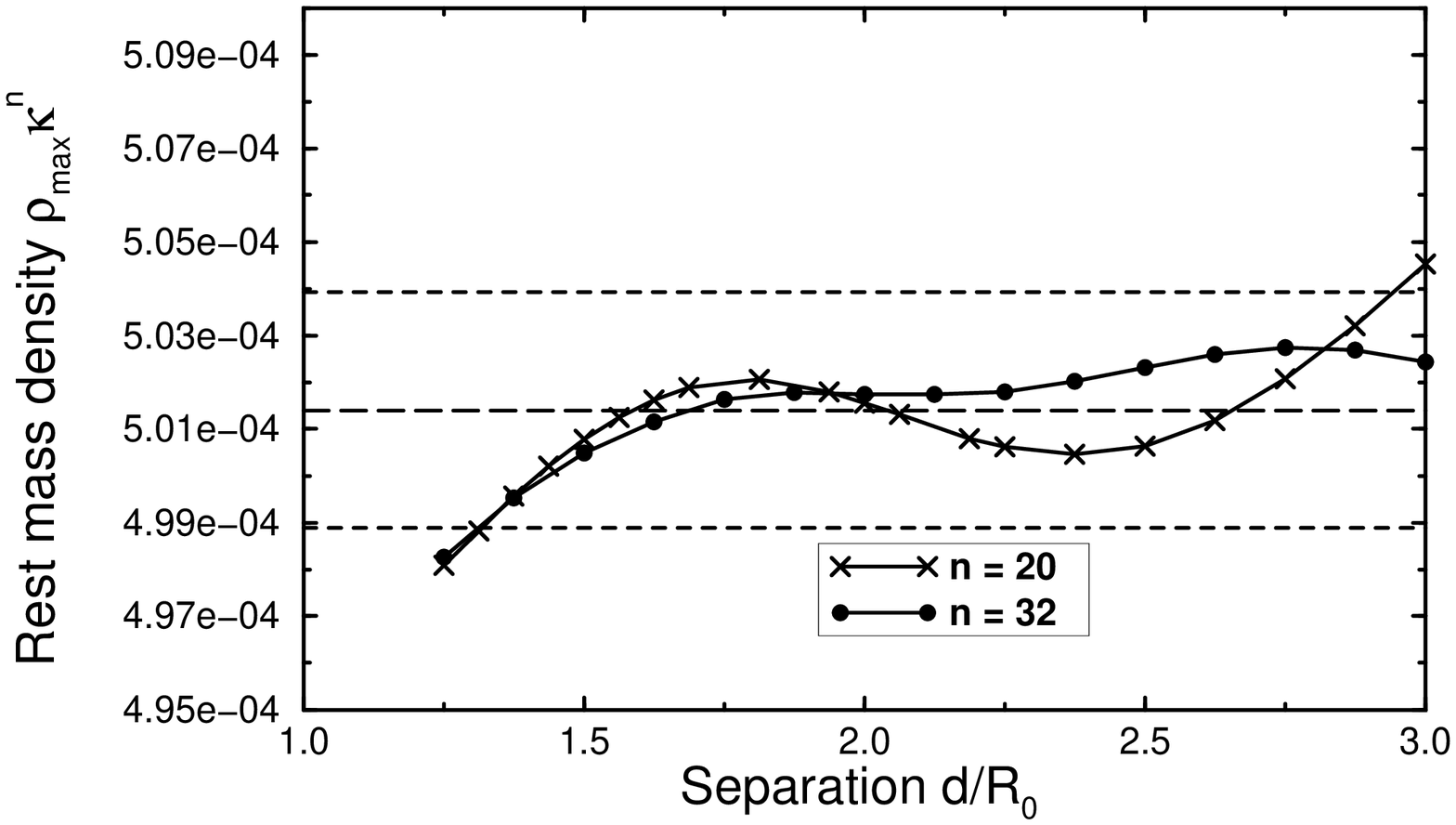}
\end{center}
\caption{The dependence of $\rho_{\rm max}(d/R_0)$ 
on the order of Legendre expansion of the gravitational
potential $n_{\rm max}$ for the case with $(M/R)_\infty=0.001$.  
Each line corresponds to a different value of $n_{\rm max}$. 
(a) The mesh type M is used for computations. 
(b) The mesh type L is used for computations. 
The horizontal long dashed line shows the $\rho_{\rm max}$ of an 
isolated spherical star $\rho_{\rm max,\infty}$.  
The horizontal short dashed lines show $\pm 0.5 \%$ difference 
from this value. 
\label{figconvrholg}}
\end{figure}

To clarify the {\it very small} non-monotonic behavior of 
$\rho_{\rm max}(d/R_0)$, we have computed sequences with 
$M/R=0.001$ by changing the number of the Legendre functions in the expansion 
for the gravitational potential, $n_{\rm max}$, in Eq.~(\ref{legen}).  
In Figure \ref{figconvrholg}(a), we plot $\rho_{\rm max}(d/R_0)$
for $n_{\rm max} = 16$, $20$ and $24$, using the mesh type M.  
For $n_{\rm max}=16$, the relative difference from the asymptotic value 
becomes $\lo 1.5 \%$ at larger separations $d/R_0 \go 2.5$. 
On the other hand, the differences between the $n_{\rm max}=20$ and 
$n_{\rm max}=24$ configurations are less than $0.5 \%$ everywhere and  
$\rho_{\rm max}(d/R_0)$ at larger separation almost reaches
the asymptotic value $\rho_{\rm max,\infty}$.  
From this figure we may conclude that the number of the Legendre 
functions in the expansion, $n_{\rm max}$, should be larger than 20.  
We also note that there arises a restriction for $n_{\rm max}$ and 
$l_{\rm max}$ from the number of grid points, since the number 
of numerical grid points should be enough to resolve the periodic 
behavior as well as  the orthogonality of the Legendre functions and 
trigonometric functions used in Eq.~(\ref{legen}) and Eq.~(\ref{sph}). 
In Figure \ref{figconvrholg}(b), we plot $\rho_{\rm max}(d/R_0)$
for the $n_{\rm max} = 20$ and $32$ sequences computed by using 
the mesh type L. The result with $n_{\rm max}=32$ shows 
that the density is almost constant for $1.8\lo d/R_0 \lo 2.2$. 
We stress again that the present numerical computation is 
targeted to compute highly deformed configurations of component 
stars with $d/R_0 \lo 2.0$.  Since all of the curves behave almost 
similarly for $d/R_0 \lo 2.0$, our computations are accurate enough 
to discuss the 
increase or decrease of $\rho_{\rm max}(d/R_0)$ of the solution 
sequence.  Therefore a reasonable choice is $n_{\rm max}\go 20$. 
It should be noted that the integrated physical quantities depend 
little on the number $n_{\rm max}$ if $n_{\rm max} \go 20$. 

Concerning the $l_{\rm max}$, that is the number of terms in the 
expansion for the velocity potential $\Phi$, this does not affect 
local or global quantities when we choose $l_{\rm max} > 8$.  
For instance, 
the choice of grid type L and $l_{\rm max} = 12$ is enough to 
compute $\Phi$ accurately at any separation.  
%
%
%

\subsection{Quasi-equilibrium sequences of irrotational binary systems 
in GR}

\begin{figure}
\epsfxsize=8.5truecm
\begin{center}
\epsffile{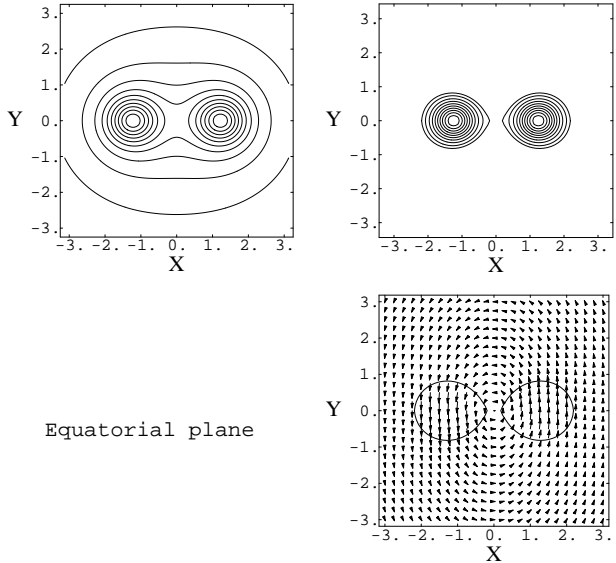}
\end{center}
\caption{Contours of the conformal factor $\Psi^4$ (upper left panel), 
the rest mass density distribution (upper right panel) and 
the shift vector with the stellar surfaces  (lower right panel), 
for the model with $(M/R)_\infty=0.14$ and $\widehat{d} = 1.25$ 
in $XY$-plane ($\theta_g = \theta_f = \pi/2$-plane). 
Lengths in the figures are normalized with $R_0$. 
\label{fcon1}}
\end{figure}

\begin{figure}
\epsfxsize=8.5truecm
\begin{center}
\epsffile{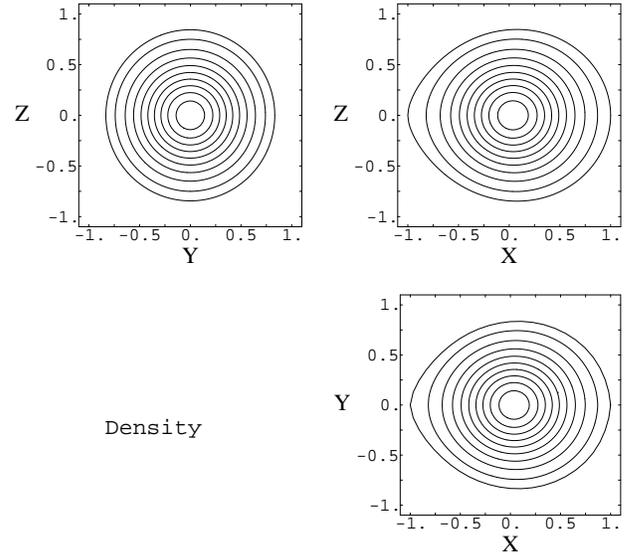}
\end{center}
\epsfxsize=8.5truecm
\begin{center}
\epsffile{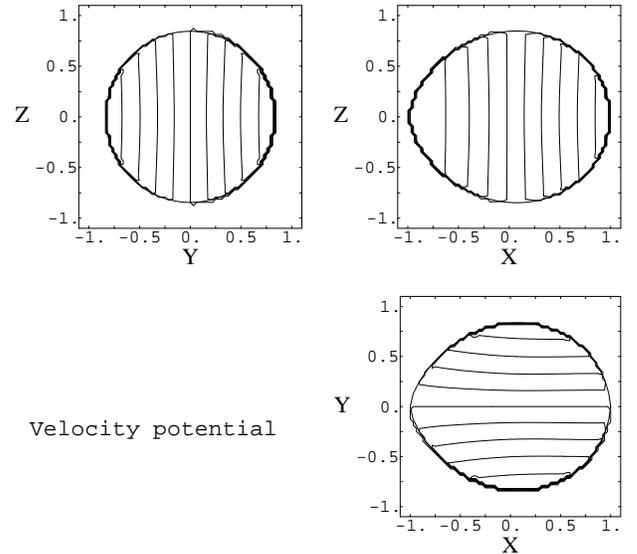}
\end{center}
\caption{Structure of a component star of an irrotational BNS system.  
Contours of the distribution of rest mass density $\rho$ 
(upper three panels) and of the velocity potential $\Phi$ 
(lower three panels) are drawn 
for the model with $(M/R)_\infty=0.17$ and $\widehat{d} = 1.25$.  
$YZ$-plane (upper left panel), $ZX$-plane (upper right) and 
$XY$-plane (lower right panel) correspond to the 
$\varphi_f=0$ and $\pi$ plane, the $\varphi_f=\pi/2$ plane, and 
the $\theta_f=\pi/2$ plane, respectively.  
Lengths in the figures are normalized with $R_0$. 
\label{fcon2}}
\end{figure}

In Figures \ref{fcon1} and \ref{fcon2} we show contours for GR 
irrotational BNS systems.  We have used mesh types L and Ls for 
the computations in this subsection.  As seen from these figures, 
our numerical method can handle highly deformed configurations of 
BNS systems and the gravitational field around them.  
In particular, in the right panel of Fig.~\ref{fcon1}, we show the 
contours for the rest mass density $\rho$ with $(M/R)_\infty = 0.14$.  
From this figure, we can see that a cusp-like structure appears at 
the inner edge of stars before the two stars make contact with 
each other \cite{errata}.  
The cusp is supposed to correspond to the inner Lagrange point 
(L1 point), with mass overflow occurring from this point when the 
separation of two stars becomes smaller enough.  


\begin{figure}
\epsfxsize=8truecm
\begin{center}
\epsffile{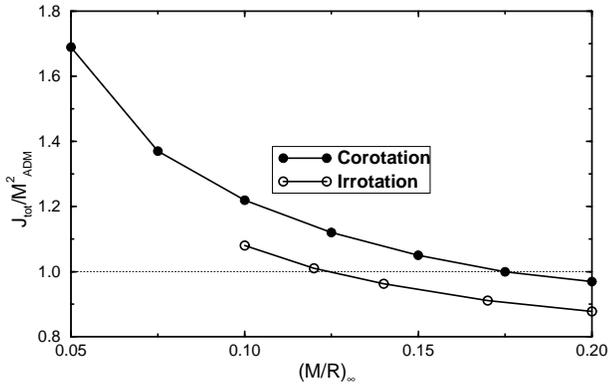}
\end{center}
\caption{Comparison of $J_{\rm tot}/M_{\rm ADM}^2$ between co-rotational 
and irrotational BNS systems at the closest distances for various 
values of the compactness parameter $(M/R)_\infty$. 
For the co-rotating models, values for the contact phases are plotted.
For the irrotational models, those for the configurations with cusps
are displayed.
\label{jom}}
\end{figure}

In Table \ref{tabjm}, we tabulate the values of physical quantities
for irrotational BNS systems with a cusp-like structure for various 
degrees of compactness.  We also plot the dimensionless total 
angular momentum $J_{\rm tot}/M_{\rm ADM}^2$ in Fig.~\ref{jom} 
which is unity for an extreme Kerr black hole.  
For the co-rotating models \cite{ba97}, $J_{\rm tot}/M_{\rm ADM}^2$ 
becomes unity around $(M/R)_\infty \sim 0.175$.  
On the other hand, this value for irrotational BNS systems becomes 
unity around $(M/R)_\infty \sim 0.12$ as shown in Table \ref{tabjm}.  
Since there exists a velocity field which rotates in the counter 
direction with respect to the orbital motion of each component star, 
the total angular momentum becomes smaller. Therefore,  after 
coalescence of the two stars due to GW emission, irrotational BNS 
systems could form a single Kerr black hole by losing much less 
angular momentum from the system than co-rotating BNS systems.  

\begin{table}
\begin{center}
\begin{tabular}{cccccc}
$(M/R)_\infty$ & $\bar M_0$ & $\bar M$ & $\Omega M_0$ & 
$J_{\rm tot}/M_{\rm ADM}^2$ & $\bar{R}_0$ \\[1mm]
\tableline
0.1  & 0.112 & 0.105 & 0.0110 & 1.08  & 1.07  \\[1mm]
0.12 & 0.130 & 0.121 & 0.0148 & 1.01  & 0.979 \\[1mm]
0.14 & 0.146 & 0.135 & 0.0192 & 0.963 & 0.899 \\[1mm]
0.17 & 0.166 & 0.150 & 0.0271 & 0.911 & 0.783 
\end{tabular}
\end{center}
\caption{Physical quantities of the irrotational BNS near 
to the configuration with a cusp-like structure (L1 point).  
\label{tabjm}}
\end{table}

\begin{figure}
\epsfxsize=8truecm
\begin{center}
\epsffile{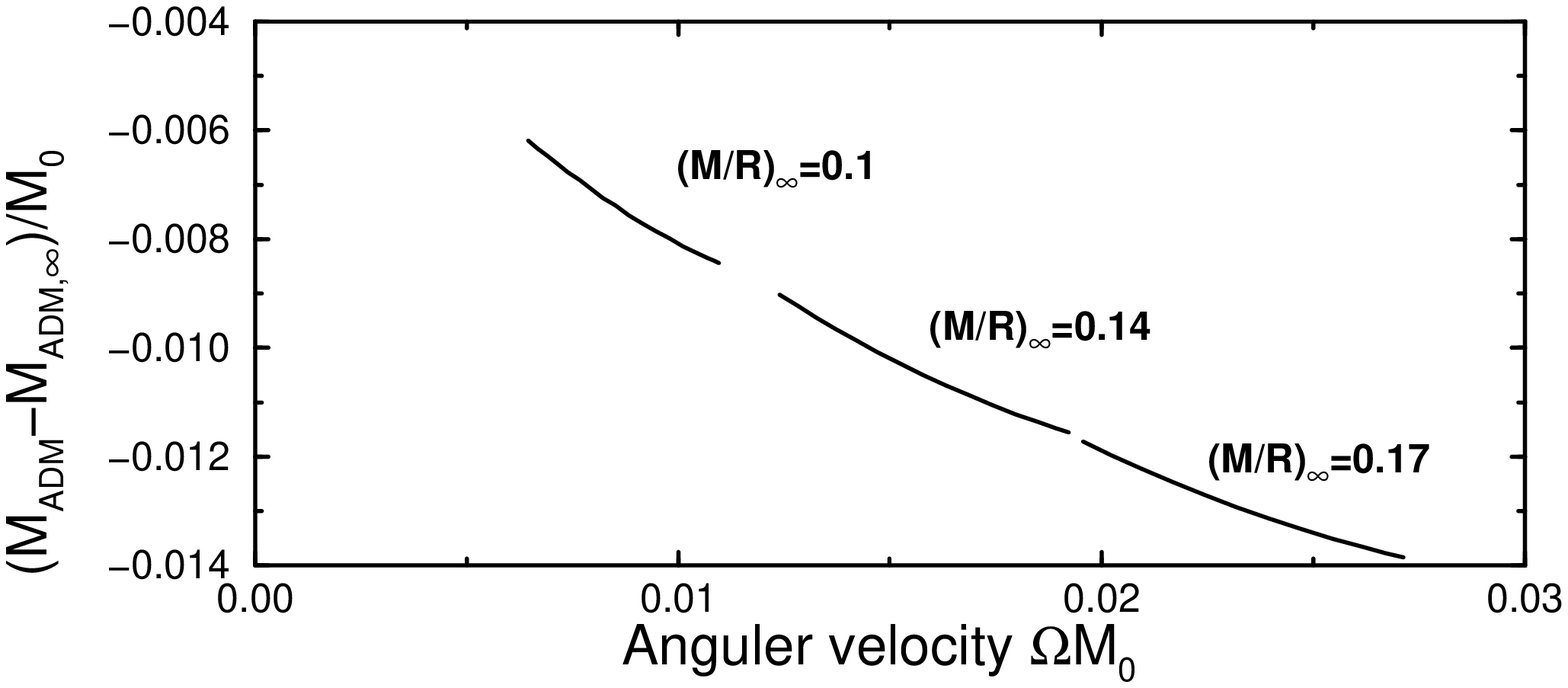}
\end{center}
\epsfxsize=8truecm
\begin{center}
\epsffile{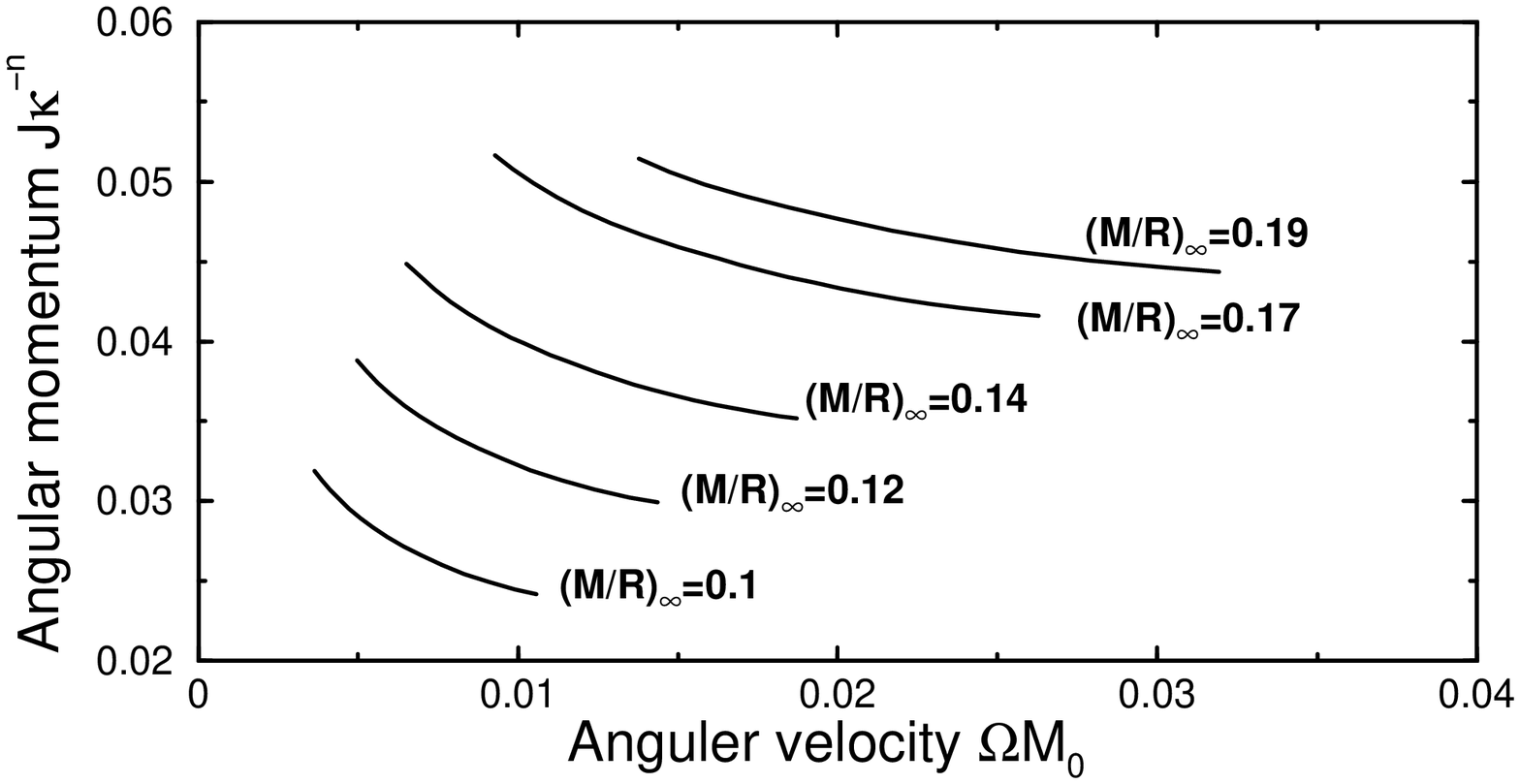}
\end{center}
\caption{The binding energy (upper panel) and the angular momentum 
(lower panel) are plotted against the orbital angular velocity.  Each 
curve corresponds to an $M_0= {\rm const}$ sequence. The value of 
the compactness $(M/R)_\infty$ of an isolated star is attached to 
each curve.  Cusp-like shape appears for the terminal model with 
the largest $\Omega M_0$ along each curve.  
\label{fseq}}
\end{figure}

In Figures \ref{fseq}, 
we show quasi-equilibrium sequences for several values of 
$(M/R)_\infty$.  
The binding energy $(M-M_\infty)/M_\infty$ and 
$\bar{J}$ are plotted against the normalized orbital angular velocity 
$\Omega M_0$.  Since each curve is a sequence with constant rest mass $M_0$, 
an evolutionary track of a BNS system as a result of GW emission 
can be approximated by following each curve from smaller to larger 
values of $\Omega M_0$.  The terminal points of the curves, i.e. 
the points with the largest $\Omega M_0$, roughly correspond to 
configurations with an L1 point.  In analogy with Newtonian models 
as well as with GR co-rotating binary systems 
\cite{ba97,lrs93}, turning points on these curves are expected to 
correspond to the points where dynamical instability sets in.  
Our results show that none of the systems with different values of 
the compactness $(M/R)_\infty$ have turning points.  
Therefore irrotational binary configurations with $n=1$ are thought to be 
dynamically stable.  In other words, at the final inspiraling phase, 
it is probable that the mass exchange starts earlier than the onset of
the orbital motion instability.  

\subsection{Individual collapse of the irrotational BNS system}
\label{sec4d}

There are two main reasons why the close binary system of irrotational 
stars has attracted wide attention.  One is that, as mentioned 
in Section \ref{intro}, it is more probable that such a situation 
is realized just before coalescence due to GW emission, since 
viscosity is not strong enough to synchronize the spin of a NS and 
the orbital motion \cite{kbc92} at this stage.  The other reason is 
that the numerical simulation of irrotational binary systems done 
by Wilson, Mathews and Marronetti \cite{wm956} showed the result that 
the maximum density of each NS increases during the inspiraling due to 
the GR effect.  They proposed a new scenario for the final inspiraling 
phase of irrotational BNS systems, namely, each component star 
of a BNS system individually collapses to form a binary black hole 
system, and after that two black holes would coalesce.  

\begin{figure}
\epsfxsize=8truecm
\begin{center}
\epsffile{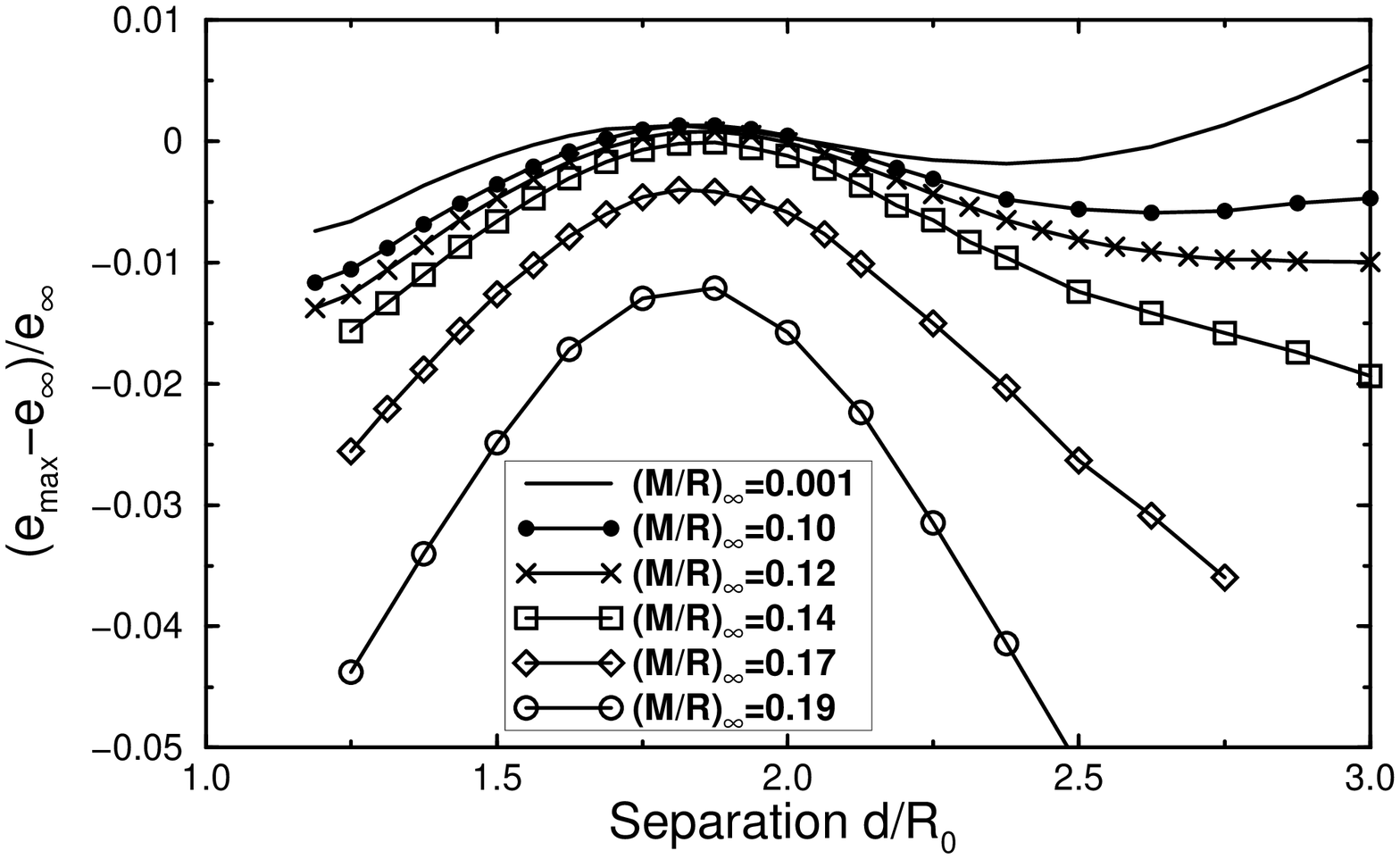}
\end{center}
\epsfxsize=8truecm
\begin{center}
\epsffile{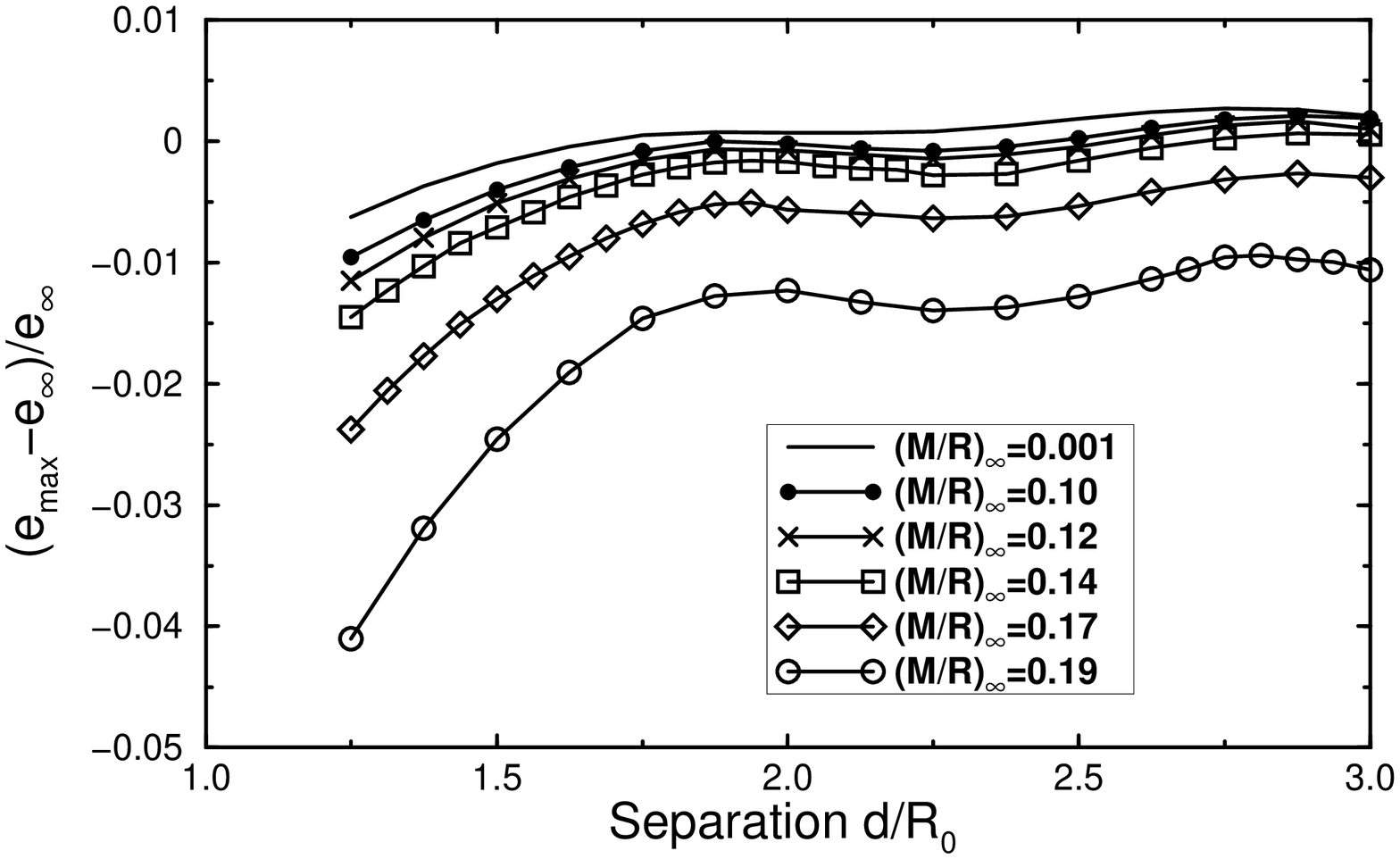}
\end{center}
\epsfxsize=8truecm
\begin{center}
\epsffile{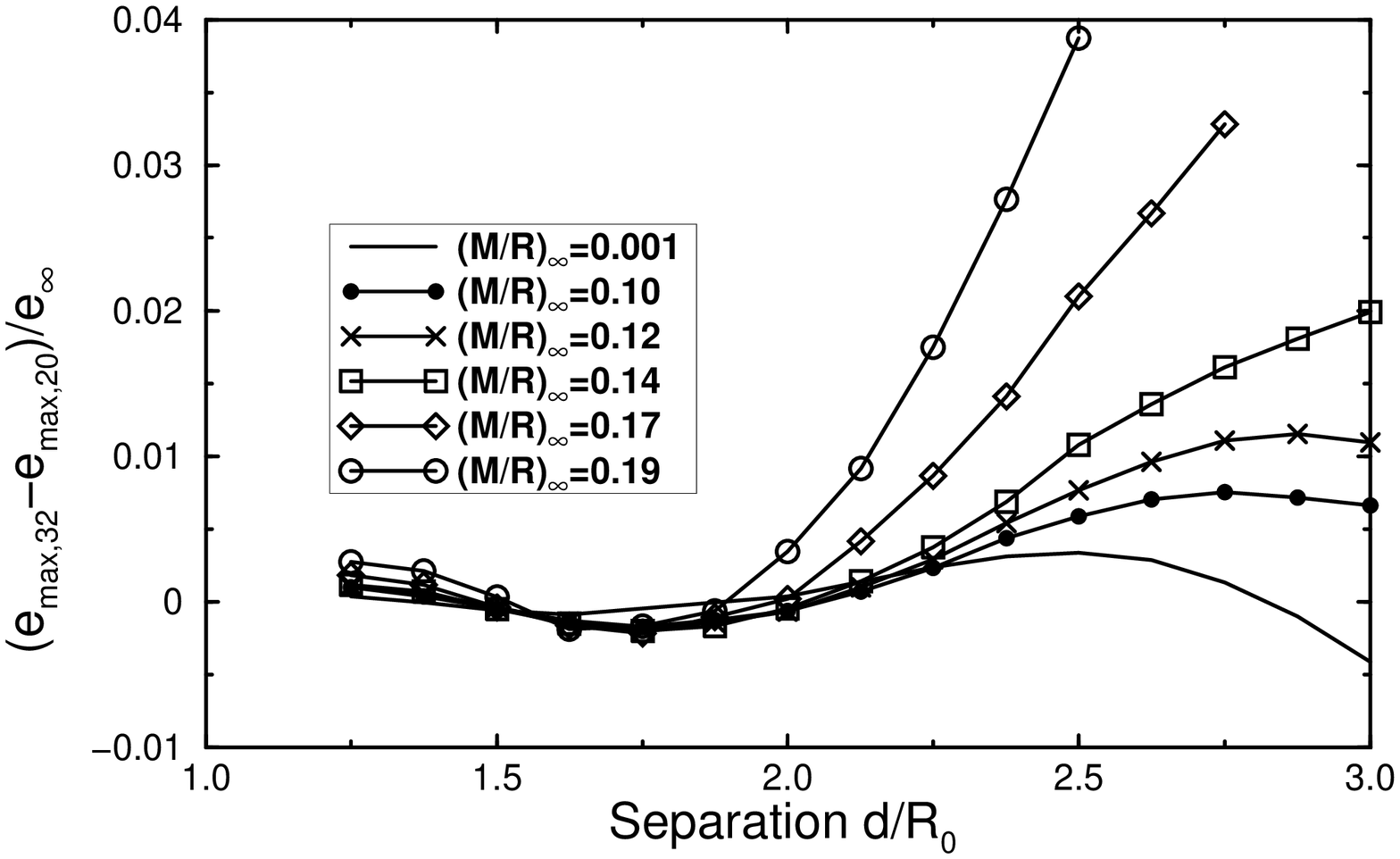}
\end{center}
\caption{
The relative change of the maximum energy density is plotted 
against the separation.  
The horizontal axis corresponds to the coordinate separation 
in units of $R_0$ for each model.  
(a) The results computed setting the order of expansion to 
$n_{\rm max} = 20$.  
Curves are plotted for $(M/R)_\infty
= 0.001$ to $0.19$.  
(b) The results computed setting the order of expansion to 
$n_{\rm max} = 32$.  Curves are plotted for $(M/R)_\infty
= 0.001$ to $0.19$.  
The maximum density does not increase as the 
separation decreases.  
(c) The differences between the results with 
$n_{\rm max} = 32$ and $n_{\rm max} = 20$.  
Curves are plotted for $(M/R)_\infty
= 0.001$ to $0.19$.  
These two results converge for $d/R_0 \lo 2.0$.  
\label{fden}}
\end{figure}

This scenario has been debated by many authors (see \cite{bgm989,mmr98} 
and references therein).  
Recent numerical computations by Marronetti, Mathews and Wilson 
\cite{mmw989} again show the maximum density increasing by a few 
percent for a sequence of models with larger compaction 
$(M/R)_\infty=0.19$, 
although the most recent re-computations by Mathews and Wilson 
give a much smaller central density increase for polytropes \cite{mw99}.  
On the other hand, the computation by the Meudon numerical relativity 
group \cite{bgm989} does not show this tendency.  In Figure \ref{fden}, 
we show our results for the relative change of the maximum energy 
density $(e_{\rm max} - e_\infty)/e_\infty$ of a star against 
the separation $\widehat d$. Here $e = \rho(1+\varepsilon)$, 
$e_{\rm max}$ is its maximum value and $e_\infty$ is its value 
in an isolated state which is computed by solving the TOV equation.  

We show the results for $n_{\rm max} = 20$ in Fig.~\ref{fden}(a) and 
$n_{\rm max} = 32$ in Fig.~\ref{fden}(b).  
We notice that the results do not depend on the order of the Legendre 
expansion $n_{\rm max}$ for $d/R_0 \lo 2.0$ with any values of 
the compactness.  
For $d/R_0 \go 2.0$, $e_{\rm max}$  does not become 
constant for $n_{\rm max} = 20$ but tends to being almost
constant when we choose $n_{\rm max} = 32$.  
To show the convergence around $d/R_0 \lo 2.0$ clearly,  
differences between the result with $n_{\rm max} = 32$ and 
$n_{\rm max} = 20$, $(e_{\rm max,\, 32} - e_{\rm max,\, 20})/e_\infty$, 
are plotted in Fig.~\ref{fden}(c).  The relative difference 
$(e_{\rm max,\, 32} - e_{\rm max,\, 20})/e_\infty$ is less than 
$0.35 \%$ at $d/R_0 \le 2.0$ for all $(M/R)_\infty \le 0.19$.  

Differences of the relative change of the central density 
$(e_{\rm max} - e_\infty)/e_\infty$ at larger separation 
$d/R_0 \go 2.0$ get larger as the compactness becomes larger.  
For instance they are $0.5\%$ and $1 \%$ for $(M/R)_\infty = 0.17$ 
and $0.19$, respectively.  
However, in any case, differences are $\lo 1 \%$ for any value of 
the compactness as shown in Fig.~\ref{fden}(b) and $e_{\rm max}$ 
tends to a constant value when the component stars are detached.  
Therefore we may discuss the change of 
$(e_{\rm max} - e_\infty)/e_\infty$ along each solution sequence.  
Our results show that the energy density changes in the range of 
$0.5 \%$ to $2.5 \%$ and decreases as the separation 
decreases.  Therefore, we do not observe the maximum density 
to increase and hence there is no tendency towards individual 
collapse in the present computation.  Also the rate of decrease of 
the maximum energy density around $d/R_0 \lo 2.0$ is larger for 
larger $(M/R)_\infty$.  

\section{Discussion and conclusion}

\subsection{Position of the present numerical method} 

As mentioned in section \ref{assump}, 
there are two different numerical schemes for 
solving irrotational binary star systems in GR \cite{bgm989,mmw989}.
The basic formulation and basic equations are almost the same but
the numerical schemes are different. 

Bonazzola, Gourgoulhon and Marck \cite{bgm989}
have used the multi-domain pseudo spectral method.
They have introduced several coordinate systems one of which
is a coordinate system fitted to the stellar surface.
The basic equations which are essentially the same as those in this
paper are solved by using the pseudo-spectral method on each of 
these domains.  Although the implementation of this numerical 
method is rather complicated, it produces numerically `exact' 
solutions with excellent accuracy.

Marronetti, Mathews and Wilson \cite{mmw989} have used
a Cartesian coordinate system and a finite difference method. 
The equations which they solved are 
also the same as ours except for their approximate treatment on the 
boundary condition for the velocity potential equation \cite{marobou}.  
Although their computation is rather less accurate, Cartesian 
coordinates are commonly used in 3D numerical relativity to compute 
dynamical coalescence of compact binary systems, 
and hence it could be advantageous for providing
initial data for future 3D simulations of BNS coalescence.  

Despite the existence of these two methods, we have presented
another scheme in this paper.  There are several reasons for this.
Firstly, since quasi-equilibrium states of binary neutron star 
systems have an important significance in GW investigations, 
it would be better to get reliable solutions by studying them 
from many directions even if this is the same problem.  
Secondly, results obtained from the two previous schemes seem 
to be somewhat different.  
In particular, the behavior of the maximum density during 
the evolution is different as discussed in subsection \ref{sec4d},
although the recent re-computation by Mathews and Wilson \cite{mw99}
gives consistent results with others for the polytropic configurations.  

In our solution method, we have used the integral form to solve Poisson 
type equations.  A spherical coordinate 
system is introduced to implement this Poisson solver.  From our 
experience with numerical computations, it seems desirable to use 
spherical coordinate systems to solve for the structures of 
component stars.  On the other hand, to solve for the gravitational 
field, it is possible to implement Cartesian coordinates and suitable 
sophisticated Poisson solvers for the coordinates (for instance, 
the multi-grid method, ICCG, and so on).  
In our scheme, the coordinates for the star are patched on those for 
the gravitational field.  This can be applied straightforwardly 
for a Cartesian grid.  
Such an implementation could also be useful for preparing initial 
conditions for numerical simulations as well as for checking
the present numerical results.  

\subsection{Future prospects}

It is important to compute quasi-equilibrium configurations 
to connect the inspiraling phase, where the perturbative 
expansion technique can be applied, to the merging phase, where 
fully dynamical computations are required.  
Such computations are also useful since the quasi-equilibrium 
solutions give accurate and appropriate initial conditions for 
simulations of BNS coalescence \cite{sh99}.  
Recently, a simulation of coalescing irrotational BNS systems in 
3D full general relativity has been performed using the solutions of 
the present paper as initial configurations \cite{su99}. 
The successful result of these simulations may be considered as 
showing that the present results are reliable for use 
as initial data for binary coalescence.  

In our formulation we have used the so called Wilson--Mathews 
formulation to treat the GR effects.  However it is known that this 
approach is exact only up to 1PN order for binary systems 
in terms of the post-Newtonian terminology, although 
2PN or higher order corrections should be included to 
express realistic BNS systems.  Asada and Shibata \cite{as967} have
formulated the Einstein equations so as to express BNS systems 
exactly up to 2PN order.  Implementing their equations in numerical 
computations may be one of the possible ways to construct 
quasi-equilibrium BNS systems, although it is necessary to solve 29 
elliptic PDEs simultaneously for the gravitational field.  Recently, 
Usui, Ury\=u and Eriguchi \cite{uue99} have developed 
a new scheme to compute quasi-equilibrium configurations by using 
a quite different form of the metric from that of the Wilson--Mathews 
formulation. 
It would be interesting to compare results from that method with our 
present results.  Also several authors have constructed 
a formulation for so called `intermediate binary black hole' (IBBH) 
states \cite{ibbh}.  This formulation could also be applicable 
to quasi-equilibrium states of BNS systems.  

\subsection{Summary and conclusion}

We have constructed a numerical method to compute quasi-equilibrium 
states of irrotational binary systems composed of compressible 
gaseous stars in general relativity.  The new numerical method is 
based on a standard finite difference method and therefore 
implementation is rather simple and straightforward.  
We have used two coordinate systems for the computation, one for the 
gravitational field variables and the other for the fluid variables.  
Physical quantities are interpolated from one to the other.  
We have shown that such a procedure works well by 
actual numerical computations of irrotational BNS systems in GR.  

The basic equation for the irrotational flow becomes an elliptic PDE 
with a Neumann type boundary condition 
at the stellar surface.  For the treatment of general relativistic 
gravity, we have used the Wilson--Mathews formulation.  In this 
formulation, the basic equations are again expressed by a system of 
elliptic PDEs.  We have developed a method to solve these PDEs.  
We have calibrated our solutions by using the results of 
independent numerical computations as well as of semi-analytic 
calculations.  We have also tested the convergence of these numerical 
solutions.  

We have succeeded in computing quasi-equilibrium sequences of 
irrotational BNS systems in GR.  As in the Newtonian case, the BNS 
systems have been found to be dynamically stable during the 
inspiraling stage until Roche lobe overflow starts at the L1 point.  
The quasi-equilibrium sequences also suggest that no substantial 
increase of the maximum energy density occurs during 
the inspiraling phase as a result of GW emission.  
Further investigations of numerical computations and implementation 
of various realistic EOS's and/or a realistic formulation of GR 
gravity are interesting and inevitable issues for theoretical 
predictions of GW signals supposed to be detected 
by the interferometric GW detectors in the next century.  

\acknowledgments

The author (KU) would like to thank 
Prof. J. C. Miller for discussions, continuous encouragement 
and helpful comments on the manuscript. 
He also would like to thank 
Prof. D. W. Sciama and Dr. A. Lanza, for their warm hospitality 
at SISSA and ICTP.  
Part of the numerical computation was carried out at the 
Astronomical Data Analysis Center of the National Astronomical 
Observatory, Japan.  

\begin{thebibliography}{99}
\bibitem{qccomment}
  Instead of this choice, we also set $q = q_c$ at the coordinate center of 
the star, that is $r_f^*=0$, in some cases.  The results do not change 
with this different choice.
\bibitem{commlg}
We also approximate a function $\chi$ in Eq.~(\ref{sph}) 
with a finite order of expansion $l_{\rm max}$. 
However this truncation does not affect this  
artificial behavior for the change of $\rho_{\rm max}$.  In fact, 
we did not observe this behavior in the purely Newtonian 
calculation \cite{ue98}, in which we used the same truncation for 
$l_{\rm max}$ but a different scheme to compute the Newtonian 
gravitational potential.
\bibitem{errata}
The cusp is formed before the two stars make contact with each other 
even for Newtonian equal mass BNS's with polytropic EOS, because of the 
non-synchronized rotation of the stars.  
This fact was overlooked in the previous paper by the present authors 
(Paper I).  Although the cusp forms just before the contact phase, 
changes of numerical results are of less than $2\%$ and the statements 
in \cite{ue98} are not affected.  The authors thank Dr. E. Gourgoulhon 
for pointing this out.
\bibitem{marobou}
We have implemented the approximate treatment on the boundary condition 
for the velocity potential equation used by \cite{mmw989}.  The result 
of $\rho_{\rm max}(d/R_0)$ was the same as the exact treatment in 
this paper for $d/R_0 \go 1.5$ but solutions did not converge for 
highly deformed configurations with $d/R_0 \lo 1.5$, 
for the case with $(M/R)_\infty = 0.17$.
\end{thebibliography}
\end{document}